\documentclass[
aps,
prd,
twocolumn,
superscriptaddress,
nofootinbib,
longbibliography,
showkeys
]{revtex4-2}

\usepackage[T1]{fontenc}
\usepackage{lmodern}
\usepackage{microtype}
\usepackage{amsmath,amssymb,amsfonts,mathtools,bm}
\usepackage{braket}
\usepackage{slashed}
\usepackage{graphicx}
\usepackage{dcolumn}
\usepackage{array}
\usepackage{booktabs}
\usepackage{multirow}
\usepackage{makecell}
\usepackage{enumitem}
\usepackage[usenames,dvipsnames,svgnames,table]{xcolor}
\usepackage[
colorlinks=true,
citecolor=Maroon,
linkcolor=MidnightBlue,
urlcolor=Maroon,
hypertexnames=true
]{hyperref}

\setlist[itemize,1]{label=$\bullet$}
\setlist[itemize,2]{label=$\circ$}

\newcommand{\dd}{\mathrm d}
\newcommand{\ee}{\mathrm e}
\newcommand{\ii}{\mathrm i}

\newcommand{\gs}{g_s}
\newcommand{\as}{\alpha'}
\newcommand{\Vcal}{\mathcal V}
\newcommand{\Wzero}{W_0}
\newcommand{\WX}{W_X}
\newcommand{\Dbar}{\overline{\rm D3}}
\newcommand{\Dthree}{{\rm D3}}
\newcommand{\mkk}{M_{\rm KK}}
\newcommand{\ms}{M_s}
\newcommand{\order}{\mathcal O}

\newcommand{\eff}{\mathrm{eff}}

\newcommand{\Schur}{\mathcal S}
\newcommand{\Aobs}{A_s}
\newcommand{\LambertW}{\operatorname{W}}
\newcommand{\Eq}[1]{Eq.~\eqref{#1}}
\newcommand{\Eqs}[2]{Eqs.~\eqref{#1}--\eqref{#2}}
\newcommand{\Sec}[1]{Sec.~\ref{#1}}
\newcommand{\Secs}[2]{Secs.~\ref{#1} and~\ref{#2}}

\newcommand{\Tab}[1]{Table~\ref{#1}}
\newcommand{\Fig}[1]{Fig.~\ref{#1}}

\newcommand{\nn}{\nonumber}

\DeclareMathOperator{\Imm}{Im}

\begin{document}
\title{The K\"ahler Stability Wall in Perturbative Brane--Antibrane Inflation}

\author{Swapnil Kumar Singh}
\email{swapnilsingh.ph@gmail.com}
\affiliation{B.M.S. College of Engineering, Bangalore, Karnataka 560019, India}

\begin{abstract}
Perturbative brane--antibrane inflation requires the physical K\"ahler volume to remain stabilized while the anti-D3 sector supplies the inflationary energy.  This paper shows that this requirement imposes an exact analytic boundary on the allowed anti-D3 tension.  In the leading nilpotent effective theory, the K\"ahler minimum and the decompactification barrier disappear through a saddle-node degeneration determined by \(U_\sigma=U_{\sigma\sigma}=0\).  The positive-tension branch of this boundary is solved in closed form and uniformized by a single real parameter.  The same solution fixes the local scaling of the K\"ahler mass, barrier height, and canonical distance to the endpoint.  Integrating out the volume modulus reveals a Schur-complement singularity, showing that the adiabatic single-field description fails at the same boundary unless the inflaton--volume mixing is suppressed.  The result is then translated into necessary flux-arithmetic filters involving the warped throat, D3 tadpole, conifold control, scalar normalization, and the full K\"ahler-sector mass matrix.  Numerical tests verify the critical locus, the saddle-node scaling laws, and the resulting sparse throat-flux filtration.  These conditions are necessary for compact realizations; no compact vacuum is asserted without the corresponding global and Wilsonian control checks.
\end{abstract}

\keywords{brane--antibrane inflation, warped throats, nilpotent superfields, K\"ahler moduli, flux compactifications, F-theory, saddle-node degenerations}

\maketitle

\tableofcontents

% ============================================================
\section{Introduction}\label{sec:intro}
% ============================================================

Brane--antibrane inflation is a geometrical mechanism for slow-roll dynamics in string theory.  The inflaton is the relative position of a mobile D3-brane and an anti-D3-brane in a compact space, and the leading large-separation interaction is the compact Green's function.  In a local approximately flat region one obtains
\begin{equation}\label{eq:intro_coulomb}
 V(r)=2T_3\left(1-\frac{1}{4\pi^2T_3r^4}+\cdots\right),
\end{equation}
where \(T_3\) is the D3-brane tension and \(r\) is the proper separation.  Warped Klebanov--Strassler regions suppress the anti-D3 tension, while Giddings--Kachru--Polchinski compactifications provide the flux background in which such throats are embedded \cite{Giddings:2001yu,Gukov:1999ya,Klebanov:1998hh,Klebanov:2000nc,Klebanov:2000hb,Grana:2005jc,DeWolfe:2002nn,Shiu:2008ry,Denef:2008wq}.  The resulting inflationary setting was developed in the original brane-inflation literature and in the KKLMMT construction \cite{Dvali:1998pa,Burgess:2001fx,Alexander:2001ks,GarciaBellido:2001ky,Kachru:2003sx,Silverstein:2003hf,Alishahiha:2004eh,Firouzjahi:2005dh,Chen:2005ad,Chen:2006nt,Lidsey:2007gq,Krause:2007jr,Pajer:2008uy,Chen:2008hz,McAllister:2008hb,Baumann:2014nda}.

The standard obstruction is the compatibility between the brane potential and K\"ahler-modulus stabilization.  In KKLMMT the non-perturbative superpotential depends on a holomorphic coordinate whose real part contains the mobile-brane position,
\begin{equation}\label{eq:intro_kklmmt_coordinate}
 T+\bar T=\Vcal^{2/3}+\gamma r^2 .
\end{equation}
Consequently the stabilized four-dimensional theory contains a Planck-suppressed inflaton mass, giving the familiar contribution \(\Delta\eta=2/3\) unless additional structure is imposed \cite{Kachru:2003sx,Kachru:2003aw,Baumann:2006th,Baumann:2007ah,Baumann:2009ni,McAllister:2007bg}.  The perturbative construction of Cicoli, Hughes, Kamal, Marino, Quevedo, Ramos-Hamud and Villa uses a different stabilization mechanism: the light volume variable during inflation is the physical volume
\begin{equation}\label{eq:sigma_def_intro}
 \sigma=\Vcal^{2/3}=T+\bar T-\frac16(\mkk r)^2,
\end{equation}
not the holomorphic real part alone \cite{Cicoli:2025origins}.  At perturbative level this removes the KKLMMT mass term and recovers the inverse-quartic brane potential in the controlled single-field limit.  The corresponding slow-roll estimate is
\begin{equation}\label{eq:intro_ns}
 n_s\simeq 1-\frac{5}{3N},\qquad r\ll1,
\end{equation}
which lies in the observationally relevant range for \(N\) of order fifty to sixty \cite{Planck:2018jri,BICEP:2021xfz,Cicoli:2025origins}.

The present paper addresses a distinct question.  Even if the inverse-quartic inflaton potential is recovered, the heavy K\"ahler-volume valley must exist at the high anti-D3 tension relevant for inflation.  The positive anti-D3 energy scales more slowly with the volume than the perturbative stabilizing terms.  Therefore, as its coefficient is increased, the local volume minimum and the adjacent decompactification barrier meet at a real codimension-one fold.  At that point the two-field effective theory loses the metastable K\"ahler trajectory about which the single-field reduction is defined.

The main result of this paper is that this fold is exactly solvable in the leading perturbative nilpotent theory.  The solution gives a closed K\"ahler-stability wall for the allowed anti-D3 tension, fixes the local saddle-node normal form, and converts the local stability condition into explicit necessary arithmetic inequalities on compact throat-flux data.  This provides an analytic pre-filter for compact searches: flux data that do not lie on the stable side of the wall cannot support the physical-volume trajectory required by the perturbative brane--antibrane scenario.

The leading two-field potential in the nilpotent branch without a linear K\"ahler term is written as
\begin{align}\label{eq:mainV_intro}
 V(\sigma,\phi)
 &=\frac{\WX(\phi)^2}{3\sigma^2}
 +3|\Wzero|^2U_{\rm pert}(\sigma),\nn\\
 U_{\rm pert}(\sigma)
 &=\frac{\alpha}{\sigma^4}
-\frac{\xi\sqrt{\gs}}{4c\,\sigma^{9/2}}
 \left(\ln\sigma-\frac{c}{\gs^2}\right).
\end{align}
where \(\phi=\sqrt{T_3}\,r\).  We define
\begin{equation}\label{eq:defs_intro}
 L\equiv\frac{c}{\gs^2},\qquad
 \beta\equiv\frac{\xi\sqrt\gs}{4c},\qquad
 a(\phi)\equiv\frac{\WX(\phi)^2}{9|\Wzero|^2},
\end{equation}
and
\begin{align}\label{eq:U_def_intro}
 V&=3|\Wzero|^2U,\nn\\
 U(\sigma,a)&=a\sigma^{-2}+\alpha\sigma^{-4}
 -\beta\sigma^{-9/2}(\ln\sigma-L).
\end{align}
The component studied in the main text is
\begin{equation}\label{eq:domain_intro}
 \sigma>0,
 \qquad
 \alpha>0,
 \qquad
 \beta>0,
 \qquad
 L>0,
 \qquad
 a\ge0 .
\end{equation}
When \(L>0\), one has \(c>0\); hence \(\beta>0\) requires the same sign as \(\xi\).  In compact type IIB orientifolds this is a topological sign condition because \(\xi=-\chi(X)\zeta(3)/[2(2\pi)^3]\) in the BBHL convention \cite{Becker:2002nn}.

The exact degeneration locus is determined by
\begin{equation}\label{eq:fold_intro}
 U_\sigma(\sigma_c,a_c)=0,
 \qquad
 U_{\sigma\sigma}(\sigma_c,a_c)=0.
\end{equation}
Solving these equations gives
\begin{equation}\label{eq:intro_ac_sigma}
 a_c(\sigma_c)=\frac{9\beta}{10}\sigma_c^{-5/2}-\frac{2\alpha}{5}\sigma_c^{-2},
\end{equation}
with
\begin{equation}\label{eq:intro_sigcrit}
 \ln\sigma_c-L=\frac{28}{45}
 +\frac{32}{45}\frac{\alpha}{\beta}\sqrt{\sigma_c}.
\end{equation}
The positive-tension branch is most simply expressed through
\begin{equation}\label{eq:y_intro}
 y\equiv \ee^{-10/9}\theta\ee^{-L/2}\sqrt{\sigma_c},
 \qquad 0<y<1,
\end{equation}
where \(\theta\) is defined in \Eq{eq:theta_def}.  One obtains the elementary parameterization
\begin{align}\label{eq:intro_uniform}
 \theta(y)&=y\exp\left[\frac45(1-y)\right],\nn\\
 \sigma_c(y)&=\exp\left[L+\frac{20}{9}-\frac85(1-y)\right],\nn\\
 a_c(y)&=\frac{9\beta}{10}\exp\left[-\frac{5L}{2}-\frac{14}{9}-4y\right](1-y).
\end{align}
The endpoint \(y\to1\) is the point at which the admissible anti-D3 tension vanishes.  The local fold coefficients are governed by the simple invariant
\begin{equation}\label{eq:Xi_intro}
 \Xi(y)=5-4y,
 \qquad 1<\Xi<5.
\end{equation}
The paper derives all of these statements from \Eq{eq:U_def_intro}; no compact vacuum is claimed from the local calculation alone.

The organization is as follows.  \Sec{sec:nilpotent_eft} derives the nilpotent effective potential and fixes normalization conventions.  \Sec{sec:critical_locus} obtains the exact K\"ahler-valley degeneration locus.  \Sec{sec:uniformization} gives the elementary uniformization and the endpoint expansion.  \Sec{sec:normal_form} derives the canonical saddle-node normal form.  \Sec{sec:schur} computes the Schur-complement singularity of the adiabatic single-field Hessian.  \Sec{sec:sigma_dependent} treats residual volume dependence in the Coulomb coefficient.  \Sec{sec:wilsonian} computes the displacement of the degeneration locus under omitted Wilsonian terms.  \Secs{sec:compact_flux}{sec:global_constraints} translate the local result into necessary compact-flux inequalities.  \Sec{sec:numerical_realization} verifies the analytic critical locus, the normal-form exponents, the Schur-complement singularity and the resulting integer throat-flux filtration in a benchmark realization.  \Sec{sec:conclusion} summarizes the results and outlines future directions.  The appendices collect derivative identities and auxiliary algebra.

% ============================================================
\section{The perturbative nilpotent effective theory}\label{sec:nilpotent_eft}
% ============================================================

We work in type IIB string theory on an orientifolded Calabi--Yau threefold \(X\), or in the weak-coupling limit of an F-theory compactification, with three-form flux
\begin{equation}\label{eq:G3_def}
 G_3=F_3-\tau_{\rm IIB}H_3,
 \qquad
 \tau_{\rm IIB}=C_0+\frac{\ii}{\gs} .
\end{equation}
The Gukov--Vafa--Witten superpotential is
\begin{equation}\label{eq:GVW}
 W_{\rm flux}=\int_XG_3\wedge\Omega,
\end{equation}
where \(\Omega\) is the holomorphic three-form \cite{Gukov:1999ya,Giddings:2001yu}.  After the complex-structure moduli and the axio-dilaton are fixed by \(D_AW_{\rm flux}=0\), the remaining superpotential parameter is
\begin{equation}\label{eq:W0_def}
 \Wzero=W_{\rm flux}\big|_{D_AW=0}.
\end{equation}
The phase of \(\Wzero\) may be absorbed in the one-modulus potential considered below; the formulas are written in terms of \(|\Wzero|\).

Throughout the paper four-dimensional reduced Planck units are used, \(M_{\rm P}=1\).  The physical-volume coordinate \(\sigma=\Vcal^{2/3}\), the reduced scalar potential \(U\), and the parameters \(a,\alpha,\beta,L\) are dimensionless in these units.  Whenever a Hessian is interpreted as a physical mass matrix, it is evaluated in a locally orthonormal frame for the scalar field-space metric.  The one-field formulae below use the leading canonical volume coordinate \(\chi=\sqrt{3/2}\ln\sigma\); corrections to this metric statement are part of the higher-order Wilsonian data discussed in Sec.~\ref{sec:wilsonian}.

A Klebanov--Strassler throat is locally characterized by integer fluxes
\begin{equation}\label{eq:flux_ints}
 \frac{1}{(2\pi)^2\as}\int_A F_3=M,
 \qquad
 \frac{1}{(2\pi)^2\as}\int_B H_3=-K,
\end{equation}
with leading warp exponent
\begin{equation}\label{eq:rho_def}
 \rho=\frac{2\pi K}{3\gs M}.
\end{equation}
The bulk scales in four-dimensional Planck units are
\begin{equation}\label{eq:bulk_scales}
 \ms=\frac{\gs^{1/4}}{\sqrt{4\pi\Vcal}},
 \qquad
 \mkk=\frac{1}{\sqrt{4\pi}\,\Vcal^{2/3}},
\end{equation}
while tip-localized string excitations are redshifted by \(\ee^{-\rho}\).  The conifold modulus scales as \(|z|\sim\exp[-2\pi K/(\gs M)]\), up to prefactors determined by the regular periods \cite{Klebanov:2000hb,Giddings:2001yu,DeWolfe:2002nn}.

The anti-D3-brane at the tip is represented in the four-dimensional theory by a nilpotent chiral multiplet \(X\) obeying \(X^2=0\) \cite{Volkov:1973ix,Rocek:1978nb,Lindstrom:1979kq,Ferrara:2013rsa,Bergshoeff:2015jxa,Garcia-Etxebarria:2015lif,DallAgata:2015zxp,Kallosh:2015nia,Kallosh:2015tea,Vercnocke:2016fbt,McDonough:2016der}.  The branch considered here is obtained from
\begin{align}\label{eq:KW_general}
 K&=-3\ln\left[f(\sigma)+(X+\bar X)g(\sigma)-X\bar Xh(\sigma)\right],\nn\\
 W&=\Wzero+X\WX(r).
\end{align}
The scalar in \(X\) is removed by the nilpotency constraint.  The following reduction fixes the normalization used in the rest of the paper.  Let
\begin{equation}\label{eq:Omega_nilpotent}
 \Omega_X=f(\sigma)+(X+\bar X)g(\sigma)-X\bar Xh(\sigma),
 \qquad K=-3\ln\Omega_X .
\end{equation}
At \(X=0\),
\begin{subequations}\label{eq:K_derivatives_nilpotent}
\begin{align}
 K_X&=-\frac{3g}{f},
 &K_\sigma&=-\frac{3f'}{f},\\
 K_{X\bar X}&=\frac{3(g^2+fh)}{f^2},
 &K_{\sigma\bar\sigma}&=\frac{3(f'^2-ff'')}{f^2},\\
 K_{\sigma\bar X}&=\frac{3(gf'-fg')}{f^2},
 &K_{X\bar\sigma}&=\frac{3(gf'-fg')}{f^2}.
\end{align}
\end{subequations}
The covariant derivatives of the superpotential \(W=\Wzero+X\WX\) are
\begin{equation}\label{eq:D_nilpotent}
 D_XW=\WX-\frac{3g}{f}\Wzero,
 \qquad
 D_\sigma W=-\frac{3f'}{f}\Wzero .
\end{equation}
The determinant of the \((\sigma,X)\) K\"ahler metric is
\begin{align}\label{eq:metric_det_nilpotent}
 \det K_{I\bar J}\big|_{X=0}
 ={}&\frac{9}{f^4}\big[
 f'^2h-ff''h+2gf'g'\nn\\
 &\hspace{3.5em}-fg'^2-f''g^2\big].
\end{align}
Substituting \eqref{eq:K_derivatives_nilpotent}--\eqref{eq:metric_det_nilpotent} into
\begin{equation}\label{eq:Fterm_general}
 V_F=\ee^K\left(K^{I\bar J}D_IW D_{\bar J}\overline W-3|W|^2\right)
\end{equation}
and then setting \(X=0\), with the phase of \(\Wzero\) chosen so that \(\Wzero\) and \(\WX\) are real along the trajectory, gives
\begin{widetext}
\begin{equation}\label{eq:Vgeneral}
 V=\frac{(f'\WX-3g'\Wzero)^2
 -f''\left(f\WX^2-6g\WX\Wzero-9h\Wzero^2\right)}
 {3f^2\left(2gf'g'-fg'^2+f'^2h-f''(g^2+fh)\right)} .
\end{equation}
\end{widetext}
Here primes denote derivatives with respect to \(\sigma\).  Setting \(g=0\) gives
\begin{equation}\label{eq:gzero_exact}
 V=\frac{f'^2\WX^2-f''(f\WX^2-9h\Wzero^2)}{3f^2(f'^2h-ff''h)} .
\end{equation}
For \(h=1\), this separates into
\begin{equation}\label{eq:gzero_split}
 V=\frac{\WX^2}{3f^2}
 +\frac{3|\Wzero|^2f''}{f^2(f'^2-ff'')} .
\end{equation}
Taking \(f(\sigma)=\sigma+\Delta f(\sigma)\), with \(|\Delta f|\ll\sigma\) in the regime in which the four-dimensional expansion is used, the anti-D3 positive-energy term is
\begin{equation}\label{eq:antiD3_leading}
 V_{\Dbar}=\frac{\WX(r)^2}{3\sigma^2}
 +\order\left(\frac{\Delta f}{\sigma}\right)\frac{\WX(r)^2}{\sigma^2}.
\end{equation}
The leading \(\sigma^{-2}\) term is therefore the one entering the exact local analysis below.  Corrections induced by \(\Delta f\) are treated as Wilsonian deformations in \Sec{sec:wilsonian}.

The nilpotent coefficient is matched to the warped anti-D3 tension and to the D3--anti-D3 Coulomb interaction.  At large separation,
\begin{align}\label{eq:WX_form}
 \WX(r)&=\ee^{-2\rho}\left[\frac{3}{4\pi}
 \left(1-\frac{\mathcal D_0}{(r\mkk)^4}\right)\right]^{1/2},\nn\\
 \mathcal D_0&=\left(\frac{3}{4\pi}\right)^3\ee^{-4\rho}.
\end{align}
With \(\phi=\sqrt{T_3}\,r\), this gives
\begin{align}\label{eq:aphi_def}
 a(\phi)&=a_0\left(1-\frac{\lambda}{\phi^4}\right),\nn\\
 a_0&=\frac{\ee^{-4\rho}}{12\pi|\Wzero|^2},\nn\\
 \lambda&=\frac{\mathcal D_0T_3^2}{\mkk^4}.
\end{align}
The constant \(a_0\) is the plateau value of the anti-D3 tension parameter in the two-field potential.

The perturbative closed-string contribution relevant for the branch without a linear K\"ahler term is
\begin{equation}\label{eq:Vpert}
 V_{\rm pert}=3|\Wzero|^2\left[
 \frac{\alpha}{\sigma^4}
 -\frac{\xi\sqrt\gs}{4c\,\sigma^{9/2}}
 \left(\ln\sigma-\frac{c}{\gs^2}\right)
 \right].
\end{equation}
The ingredients entering this expression arise from the BBHL correction, string loops, logarithmic field redefinitions and higher-derivative curvature terms in perturbative K\"ahler stabilization \cite{Becker:2002nn,Berg:2005ja,vonGersdorff:2005bf,Berg:2007wt,Cicoli:2007xp,Cicoli:2008gp,Conlon:2008cj,Antoniadis:2018hqy,Antoniadis:2019rkh,Grimm:2017pid,Grimm:2017okk,Cicoli:2023opf,Cicoli:2024fva}.  Combining \Eqs{eq:antiD3_leading}{eq:Vpert} yields \Eq{eq:U_def_intro}.  This is the potential analyzed in the rest of the paper.

% ============================================================
\section{Critical anti-D3 tension and the K\"ahler-valley degeneration locus}\label{sec:critical_locus}
% ============================================================

The analytic structure is clearest if one begins with a general affine deformation of a one-dimensional reduced scalar potential,
\begin{equation}\label{eq:general_affine}
 U(\sigma,A)=Aw(\sigma)+P(\sigma),
\end{equation}
where \(A\) is an external positive-energy coefficient and \(w'(\sigma)\neq0\) along the component under consideration.  A fold of stationary points obeys
\begin{equation}\label{eq:general_fold}
 A_cw'(\sigma_c)+P'(\sigma_c)=0,
 \qquad
 A_cw''(\sigma_c)+P''(\sigma_c)=0.
\end{equation}
Eliminating \(A_c\) gives the intrinsic critical equation
\begin{equation}\label{eq:general_fold_operator}
 P''(\sigma_c)w'(\sigma_c)-P'(\sigma_c)w''(\sigma_c)=0,
\end{equation}
and the corresponding value of the external coefficient is
\begin{equation}\label{eq:general_Ac}
 A_c(\sigma_c)=-\frac{P'(\sigma_c)}{w'(\sigma_c)}.
\end{equation}
The singularity is a genuine saddle-node whenever
\begin{equation}\label{eq:cubic_nonzero_general}
 U_{\sigma\sigma\sigma,c}=A_cw'''(\sigma_c)+P'''(\sigma_c)\neq0.
\end{equation}

For the string-theory potential \Eq{eq:U_def_intro},
\begin{equation}\label{eq:wP_main}
 w(\sigma)=\sigma^{-2},
 \qquad
 P(\sigma)=\alpha\sigma^{-4}-\beta\sigma^{-9/2}(\ln\sigma-L).
\end{equation}
Therefore \Eq{eq:general_fold_operator} reduces to
\begin{equation}\label{eq:sigmaP_equation}
 \sigma P''(\sigma_c)+3P'(\sigma_c)=0,
\end{equation}
while \Eq{eq:general_Ac} gives
\begin{equation}\label{eq:ac_fromP}
 a_c(\sigma_c)=\frac12\sigma_c^3P'(\sigma_c).
\end{equation}

For later use we derive the same result in terms of first-order functions.  Differentiating \Eq{eq:U_def_intro} gives
\begin{align}\label{eq:Usigma}
 U_\sigma={}&-2a\sigma^{-3}-4\alpha\sigma^{-5}
 +\beta\sigma^{-11/2}
 \left[\frac92(\ln\sigma-L)-1\right]
 \nn\\
={}&\sigma^{-11/2}F(\sigma,a),
\end{align}
where
\begin{equation}\label{eq:Fdef}
 F(\sigma,a)=\beta\left[\frac92(\ln\sigma-L)-1\right]
 -4\alpha\sigma^{1/2}-2a\sigma^{5/2}.
\end{equation}
The second derivative is
\begin{equation}\label{eq:Usecond}
 U_{\sigma\sigma}=-\frac{11}{2}\sigma^{-13/2}F
 +\sigma^{-11/2}F_\sigma,
\end{equation}
with
\begin{equation}\label{eq:Fsigma}
 F_\sigma=\frac{9\beta}{2\sigma}-2\alpha\sigma^{-1/2}-5a\sigma^{3/2}.
\end{equation}
At a stationary point, \(F=0\), so the sign of the \(\sigma\)-curvature is the sign of \(F_\sigma\).

The degeneration point satisfies
\begin{equation}\label{eq:foldconds}
 F(\sigma_c,a_c)=0,
 \qquad
 F_\sigma(\sigma_c,a_c)=0.
\end{equation}
Solving the second equation for \(a_c\) gives
\begin{equation}\label{eq:acrit_sigma}
 a_c(\sigma_c)=\frac{9\beta}{10}\sigma_c^{-5/2}
 -\frac{2\alpha}{5}\sigma_c^{-2}.
\end{equation}
Substituting into \(F=0\) gives
\begin{equation}\label{eq:sigcrit}
 \ln\sigma_c-L=\frac{28}{45}
 +\frac{32}{45}\frac{\alpha}{\beta}\sqrt{\sigma_c}.
\end{equation}
Positive anti-D3 tension at the degeneration point requires
\begin{equation}\label{eq:positive_ac}
 a_c>0
 \quad\Longleftrightarrow\quad
 \sqrt{\sigma_c}<\frac{9\beta}{4\alpha}.
\end{equation}

The nature of the two extrema follows directly from the next derivative.  One has
\begin{align}\label{eq:F_higher}
 F_{\sigma\sigma}&=-\frac{9\beta}{2\sigma^2}
 +\alpha\sigma^{-3/2}
 -\frac{15}{2}a\sigma^{1/2},\nn\\
 F_{\sigma\sigma\sigma}&=\frac{9\beta}{\sigma^3}
 -\frac32\alpha\sigma^{-5/2}
 -\frac{15}{4}a\sigma^{-1/2}.
\end{align}
Using \Eq{eq:acrit_sigma},
\begin{equation}\label{eq:Fss_c_general}
 F_{\sigma\sigma,c}=\sigma_c^{-2}
 \left(4\alpha\sqrt{\sigma_c}-\frac{45}{4}\beta\right).
\end{equation}
The positivity condition \Eq{eq:positive_ac} implies
\begin{equation}\label{eq:Fss_negative}
 F_{\sigma\sigma,c}<0.
\end{equation}
The lower-\(\sigma\) root is the local minimum and the higher-\(\sigma\) root is the decompactification barrier, because near the fold \(F_\sigma\simeq F_{\sigma\sigma,c}(\sigma-\sigma_c)\).

% ============================================================
\section{Elementary uniformization of the positive-tension branch}\label{sec:uniformization}
% ============================================================

The critical equation \Eq{eq:sigcrit} admits a Lambert representation.  It also admits a simpler elementary uniformization on the positive-tension component.  Define
\begin{equation}\label{eq:theta_def}
 \theta=\left(\frac{16c\ee^{10/9}}{9\xi\sqrt\gs}\right)\alpha\ee^{L/2},
 \qquad
 u=\ee^{-L/2}\sqrt{\sigma_c}.
\end{equation}
Since \(\beta=\xi\sqrt\gs/(4c)\), this is equivalent to
\begin{equation}\label{eq:alpha_beta_theta}
 \frac{\alpha\ee^{L/2}}{\beta}=\frac94\ee^{-10/9}\theta .
\end{equation}
Equation \eqref{eq:sigcrit} becomes
\begin{equation}\label{eq:u_eq}
 2\ln u=\frac{28}{45}+\frac85\ee^{-10/9}\theta u.
\end{equation}
Solving for \(u\) gives
\begin{equation}\label{eq:Lambert_u}
 u_k(\theta)=-\frac54\frac{\ee^{10/9}}{\theta}
 \LambertW_k\left[-\frac45\theta\ee^{-4/5}\right].
\end{equation}
For \(0<\theta<1\), the positive-tension component is the principal branch \(k=0\).

Now define
\begin{equation}\label{eq:y_def}
 y=\ee^{-10/9}\theta u.
\end{equation}
Then \Eq{eq:u_eq} becomes
\begin{equation}\label{eq:y_equation}
 2\ln\left(\frac{y}{\theta}\right)+\frac85=\frac85y.
\end{equation}
Solving for \(\theta\) gives
\begin{subequations}\label{eq:uniform_branch}
\begin{align}
 \theta(y)&=y\exp\left[\frac45(1-y)\right],\label{eq:theta_y}\\
 \sigma_c(y)&=\exp\left[L+\frac{20}{9}-\frac85(1-y)\right],\label{eq:sigma_y}\\
 \alpha(y)&=\frac{9\beta}{4}\,y\,\sigma_c(y)^{-1/2},\label{eq:alpha_y}\\
 a_c(y)&=\frac{9\beta}{10}\exp\left[-\frac{5L}{2}-\frac{14}{9}-4y\right](1-y).
 \label{eq:ac_y}
\end{align}
\end{subequations}
The range is
\begin{equation}\label{eq:y_range}
 0<y<1.
\end{equation}
Indeed,
\begin{equation}\label{eq:theta_monotone}
 \frac{\dd\ln\theta}{\dd y}=\frac{1}{y}-\frac45>0
\end{equation}
throughout this interval, so \(y\) is a global coordinate on the positive-tension component.  The point \(y\to1\) is the endpoint \(\theta\to1\), where \(a_c\to0\).  For a fixed effective theory, \(\theta\) is fixed by \((\alpha,\beta,L)\), and \(y\) is therefore determined by \Eq{eq:theta_y}.  Thus \(y\) is not an additional modulus.  It is a coordinate on the family of critical points obtained as the microscopic perturbative data are varied, or equivalently on the image of a compact-flux scan after the K\"ahler reduction has produced \((\alpha,\beta,L)\).

The endpoint expansion is useful for estimating the sensitivity of the construction.  Set
\begin{equation}\label{eq:s_delta_def}
 s=1-y,
 \qquad
 \delta=1-\theta.
\end{equation}
Then
\begin{equation}\label{eq:theta_s_exp}
 \theta=(1-s)\ee^{4s/5}
 =1-\frac{s}{5}-\frac{12s^2}{25}-\frac{88s^3}{375}+\order(s^4),
\end{equation}
so \(s=5\delta+\order(\delta^2)\).  Substituting into \Eq{eq:ac_y} gives
\begin{equation}\label{eq:ac_delta}
 a_c=\frac{9\beta}{2}\ee^{-5L/2-50/9}\delta
 \left[1+8\delta-\frac{64}{3}\delta^2+\order(\delta^3)\right].
\end{equation}
The critical anti-D3 tension therefore vanishes linearly at the endpoint.

The late-time zero-tension extrema have a different analytic structure.  Setting \(a=0\) in \(F=0\) gives
\begin{equation}\label{eq:late_extremum}
 \beta\left[\frac92(\ln\sigma-L)-1\right]-4\alpha\sqrt\sigma=0.
\end{equation}
With \(x=\ee^{-L/2}\sqrt\sigma\),
\begin{equation}\label{eq:late_x_eq}
 \ln x=\frac19+\ee^{-10/9}\theta x,
\end{equation}
so
\begin{equation}\label{eq:late_W}
 x_k=-\frac{\ee^{10/9}}{\theta}\LambertW_k\left(-\frac{\theta}{\ee}\right),
 \qquad k=0,-1.
\end{equation}
The two real branches merge at the Lambert branch point, where
\begin{equation}\label{eq:late_sigma_fold}
 \sigma_{\rm L}=\ee^{L+20/9}.
\end{equation}
Thus the late-time zero-tension merger is a square-root Lambert collision, while the positive-tension endpoint described by \Eq{eq:ac_delta} is a linear collapse in \(1-y\).

The relation to throat fluxes follows by comparing \(a_0\) with \(a_c(y)\).  The plateau value \Eq{eq:aphi_def} must obey
\begin{equation}\label{eq:a0_bound}
 a_0<a_c(y).
\end{equation}
Equivalently,
\begin{subequations}\label{eq:rho_crit_y}
\begin{align}
 \rho&>\rho_{\min}(y),\label{eq:rho_bound_y}\\
 \rho_{\min}(y)&\equiv\max\{0,\rho_c(y)\},\label{eq:rho_min_y}\\
 \rho_c(y)&=\frac{5L}{8}+y+\frac{7}{18}
 -\frac14\ln\left[\frac{54\pi}{5}\beta|\Wzero|^2(1-y)\right].
 \label{eq:rho_c_y}
\end{align}
\end{subequations}
The maximum with zero only records the fact that a physical throat already has \(\rho>0\).  When \(\rho_c(y)\le0\), the critical-tension condition does not impose an additional lower bound on the warp exponent.  Using \Eq{eq:rho_def}, the nontrivial flux-ratio condition is
\begin{equation}\label{eq:flux_ratio_bound}
 \frac{K}{M}>\frac{3\gs}{2\pi}\rho_{\min}(y).
\end{equation}
This is a necessary local condition for the K\"ahler valley to exist at the anti-D3 tension required during inflation.

% ============================================================
\section{Canonical saddle-node normal form}\label{sec:normal_form}
% ============================================================

Let
\begin{equation}\label{eq:local_vars}
 q=\sigma-\sigma_c,
 \qquad
 \mu=a_c-a .
\end{equation}
The metastable side is \(\mu>0\).  Expanding \(F\) around the degeneration point gives
\begin{align}\label{eq:F_expand}
 F={}&2\sigma_c^{5/2}\mu
 +\frac12F_{\sigma\sigma,c}q^2
 +F_{\sigma a,c}q(a-a_c)
 \nn\\
 &+\frac16F_{\sigma\sigma\sigma,c}q^3
 +\order(q^4,\mu q^2),
\end{align}
where
\begin{equation}\label{eq:Fa_derivs}
 F_a=-2\sigma^{5/2},
 \qquad
 F_{\sigma a}=-5\sigma^{3/2}.
\end{equation}
The two nearby stationary points are
\begin{equation}\label{eq:q_pm}
 q_\pm=\pm\left[-\frac{4\sigma_c^{5/2}}{F_{\sigma\sigma,c}}\mu\right]^{1/2}
 +\order(\mu).
\end{equation}
Because \(F_{\sigma\sigma,c}<0\), the root \(q_-\) is the stable local minimum and \(q_+\) is the barrier.

The next coefficient follows by writing
\begin{equation}\label{eq:q_series}
 q=s_1\mu^{1/2}+s_2\mu+s_3\mu^{3/2}+\cdots .
\end{equation}
The leading coefficient is
\begin{equation}\label{eq:s1}
 s_1=\pm\left[-\frac{4\sigma_c^{5/2}}{F_{\sigma\sigma,c}}\right]^{1/2},
\end{equation}
and the midpoint shift is
\begin{equation}\label{eq:s2}
 s_2=\frac{F_{\sigma a,c}}{F_{\sigma\sigma,c}}
 +\frac{2\sigma_c^{5/2}F_{\sigma\sigma\sigma,c}}{3F_{\sigma\sigma,c}^2}.
\end{equation}
Thus the separation begins at order \(\mu^{1/2}\), while the center of the pair moves at order \(\mu\).

The leading canonical volume field is
\begin{equation}\label{eq:chi_def}
 \chi=\sqrt{\frac32}\ln\sigma .
\end{equation}
Hence the canonical distance between the local minimum and the barrier is
\begin{align}\label{eq:Delta_chi_general}
 \Delta\chi
 &=\sqrt{\frac32}\ln\frac{\sigma_c+q_+}{\sigma_c+q_-}
 \nn\\
 &=2\sqrt6\left[-\frac{\sigma_c^{1/2}}{F_{\sigma\sigma,c}}\right]^{1/2}\mu^{1/2}
 +\order(\mu^{3/2}).
\end{align}
The canonical K\"ahler mass is
\begin{equation}\label{eq:mchi_def}
 m_\chi^2=\left(\frac{\dd\sigma}{\dd\chi}\right)^2V_{\sigma\sigma}
 =\frac23\sigma^2V_{\sigma\sigma}.
\end{equation}
At a stationary point,
\begin{equation}\label{eq:mchi_F}
 m_\chi^2=2|\Wzero|^2\sigma^{-7/2}F_\sigma.
\end{equation}
At the stable root this gives
\begin{equation}\label{eq:mchi_scaling}
 m_\chi^2=4|\Wzero|^2\sigma_c^{-7/2}
 \left(-\sigma_c^{5/2}F_{\sigma\sigma,c}\right)^{1/2}\mu^{1/2}
 +\order(\mu).
\end{equation}

The barrier height follows from
\begin{equation}\label{eq:Vsigma_F}
 V_\sigma=3|\Wzero|^2\sigma^{-11/2}F.
\end{equation}
To leading order,
\begin{align}\label{eq:Vlocal_q}
 V(q)={}&V_c+3|\Wzero|^2\sigma_c^{-11/2}
 \left[2\sigma_c^{5/2}\mu q+\frac{F_{\sigma\sigma,c}}{6}q^3\right]
 \nn\\
 &+\order(\mu q^2,q^4).
\end{align}
Therefore
\begin{equation}\label{eq:barrier_height_general}
 \Delta V_B=\frac{16|\Wzero|^2\sigma_c^{-7/4}}{\sqrt{-F_{\sigma\sigma,c}}}\mu^{3/2}
 +\order(\mu^2).
\end{equation}
The universal hierarchy is
\begin{equation}\label{eq:fold_scaling}
 \Delta\chi\sim\mu^{1/2},
 \qquad
 m_\chi^2\sim\mu^{1/2},
 \qquad
 \Delta V_B\sim\mu^{3/2}.
\end{equation}

The canonical cubic coefficient at the degeneration locus is
\begin{equation}\label{eq:cubic_chi}
 V_{\chi\chi\chi,c}=2\sqrt{\frac23}|\Wzero|^2\sigma_c^{-5/2}F_{\sigma\sigma,c}.
\end{equation}
Consequently the local potential takes the normal form
\begin{align}\label{eq:canonical_normal_form}
 V(\chi)={}&V_c-3|\Wzero|^2\sigma_c^{-2}\mu
 +2\sqrt6|\Wzero|^2\sigma_c^{-2}\mu\,\delta\chi
 \nn\\
 &+\frac16V_{\chi\chi\chi,c}\delta\chi^3
 +\order(\mu\delta\chi^2,\delta\chi^4),
\end{align}
where \(\delta\chi=\chi-\chi_c\).

On the elementary branch, \Eq{eq:alpha_y} gives
\begin{equation}\label{eq:alpha_beta_y}
 \frac{\alpha\sqrt{\sigma_c}}{\beta}=\frac94y.
\end{equation}
Thus
\begin{subequations}\label{eq:F_derivs_uniform}
\begin{align}
 F_{\sigma\sigma,c}&=-\frac{9\beta}{4}\sigma_c^{-2}(5-4y),\label{eq:Fss_uniform}\\
 F_{\sigma\sigma\sigma,c}&=\frac{45\beta}{8}\sigma_c^{-3}.
 \label{eq:Fsss_uniform}
\end{align}
\end{subequations}
Defining
\begin{equation}\label{eq:Xi_def}
 \Xi(y)=5-4y,
 \qquad 1<\Xi<5,
\end{equation}
one obtains
\begin{subequations}\label{eq:local_coeff_uniform}
\begin{align}
 \Delta\chi&=\frac{4\sqrt6}{3\sqrt{\beta\Xi}}
 \sigma_c^{5/4}\mu^{1/2}+\order(\mu^{3/2}),\label{eq:Dchi_uniform}\\
 m_\chi^2&=6|\Wzero|^2\sqrt{\beta\Xi}\,\sigma_c^{-13/4}\mu^{1/2}
 +\order(\mu),\label{eq:mchi_uniform}\\
 \Delta V_B&=\frac{32}{3\sqrt{\beta\Xi}}
 |\Wzero|^2\sigma_c^{-3/4}\mu^{3/2}+\order(\mu^2).
 \label{eq:DV_uniform}
\end{align}
\end{subequations}
These are purely local K\"ahler-sector coefficients.  Inflaton-dependent residues require the derivatives of \(a(\phi)\), as shown next.

% ============================================================
\section{Schur complement and the boundary of the adiabatic single-field description}\label{sec:schur}
% ============================================================

Let \(\Phi^I\) denote the scalar coordinates and let \(G_{IJ}(\Phi)\) be the field-space metric obtained from the four-dimensional K\"ahler potential.  The physical mass matrix is
\begin{equation}\label{eq:covariant_mass_matrix}
 {\cal M}_{IJ}=\nabla_I\nabla_JV,
 \qquad
 {\cal M}^{I}{}_{J}=G^{IK}\nabla_K\nabla_JV,
\end{equation}
where \(\nabla_I\) is the Levi-Civita connection of \(G_{IJ}\).  In the local analysis below we work in the orthonormal frame \((\phi,\chi)\) at the point of interest, with \(\chi=\sqrt{3/2}\ln\sigma\) at leading order and with \(\phi\) chosen as the canonically normalized brane coordinate.  In such a frame the connection terms vanish at the point, and the Hessian entries are the ordinary second derivatives displayed below.  Away from this local frame the same formulae are to be read as Schur complements of the covariant mass matrix \eqref{eq:covariant_mass_matrix}.

Let \(\sigma_\star(\phi)\) denote the stable solution of
\begin{equation}\label{eq:valley_eq}
 V_\sigma(\sigma_\star(\phi),\phi)=0.
\end{equation}
The adiabatic potential is
\begin{equation}\label{eq:Veff_def}
 V_{\eff}(\phi)=V(\sigma_\star(\phi),\phi).
\end{equation}
Since \(V_\sigma=0\) along the trajectory,
\begin{equation}\label{eq:Veff_first}
 \frac{\dd V_{\eff}}{\dd\phi}=V_\phi.
\end{equation}
Differentiating \Eq{eq:valley_eq} gives
\begin{equation}\label{eq:sigma_star_prime}
 \sigma_\star'=-\frac{V_{\phi\sigma}}{V_{\sigma\sigma}},
\end{equation}
and hence
\begin{equation}\label{eq:Veff_second}
 \frac{\dd^2V_{\eff}}{\dd\phi^2}
 =V_{\phi\phi}-\frac{V_{\phi\sigma}^2}{V_{\sigma\sigma}}.
\end{equation}
The second term is the Schur complement of the heavy-field Hessian.  The same structure controls heavy-field effects in multifield inflation and in the effective theory of fluctuations \cite{Gordon:2000hv,Langlois:2008qf,Tolley:2009fg,Achucarro:2010da,Peterson:2010np,Gong:2011uw,Achucarro:2012sm,Cespedes:2012hu,Burgess:2012dz,RenauxPetel:2015mga,Cheung:2007st,Weinberg:2008hq,Senatore:2010wk,Shiu:2011qw,Komatsu:2009kd,Chen:2010xka}.

For \(a=a(\phi)\),
\begin{subequations}\label{eq:phi_derivs}
\begin{align}
 V_\phi&=3|\Wzero|^2a_\phi\sigma^{-2},\label{eq:Vphi}\\
 V_{\phi\sigma}&=-6|\Wzero|^2a_\phi\sigma^{-3},\label{eq:Vphisigma}\\
 V_{\phi\phi}&=3|\Wzero|^2a_{\phi\phi}\sigma^{-2}.
 \label{eq:Vphiphi}
\end{align}
\end{subequations}
Since \(V_{\sigma\sigma}=(3/2)m_\chi^2/\sigma^2\), the Schur term is
\begin{equation}\label{eq:DeltaVpp}
 \Delta V_{\eff}''\equiv-\frac{V_{\phi\sigma}^2}{V_{\sigma\sigma}}
 =-24|\Wzero|^4\frac{a_\phi^2}{m_\chi^2}\sigma^{-4}.
\end{equation}
Using \Eq{eq:mchi_scaling},
\begin{equation}\label{eq:Schur_residue_general}
 \Delta V_{\eff}''
 =-\frac{6|\Wzero|^2a_{\phi,c}^2}{\sigma_c^{7/4}\sqrt{-F_{\sigma\sigma,c}}}
 \mu^{-1/2}+\order(\mu^0).
\end{equation}
On the elementary branch this becomes
\begin{equation}\label{eq:Schur_residue_uniform}
 \Delta V_{\eff}''
 =-\frac{4|\Wzero|^2a_{\phi,c}^2}{\sqrt{\beta\Xi(y)}\,\sigma_c^{3/4}}
 \mu^{-1/2}+\order(\mu^0).
\end{equation}
The pole is absent only if \(a_{\phi,c}\) vanishes rapidly enough to compensate the collapse of \(V_{\sigma\sigma}\).

The same expression defines a pre-degeneration control scale.  Let \(\Lambda_\phi^2\) be the curvature scale beyond which the adiabatic expansion is no longer trusted.  The Schur term reaches this scale at
\begin{equation}\label{eq:mu_schur_general}
 \mu_{\Schur}=\left[
 \frac{6|\Wzero|^2a_{\phi,c}^2}
 {\sigma_c^{7/4}\sqrt{-F_{\sigma\sigma,c}}\,\Lambda_\phi^2}
 \right]^2.
\end{equation}
For the intrinsic comparison \(\Lambda_\phi^2=|V_{\phi\phi,c}|=3|\Wzero|^2|a_{\phi\phi,c}|\sigma_c^{-2}\), one obtains
\begin{equation}\label{eq:mu_schur_intrinsic}
 \mu_{\Schur}=\left[
 \frac{2a_{\phi,c}^2\sigma_c^{1/4}}
 {|a_{\phi\phi,c}|\sqrt{-F_{\sigma\sigma,c}}}
 \right]^2,
 \qquad a_{\phi\phi,c}\neq0.
\end{equation}
The trajectory bending has the same singular exponent:
\begin{equation}\label{eq:valley_bending}
 \sigma_\star'=\frac{4|\Wzero|^2a_\phi}{\sigma m_\chi^2}
 =\frac{a_{\phi,c}\sigma_c^{5/2}}
 {\left(-\sigma_c^{5/2}F_{\sigma\sigma,c}\right)^{1/2}}
 \mu^{-1/2}+\order(\mu^0).
\end{equation}

The result is coordinate independent.  With \(\chi=\sqrt{3/2}\ln\sigma\),
\begin{equation}\label{eq:canonical_mixing}
 V_{\phi\chi}=\frac{\dd\sigma}{\dd\chi}V_{\phi\sigma},
 \qquad
 V_{\chi\chi}=\left(\frac{\dd\sigma}{\dd\chi}\right)^2V_{\sigma\sigma},
\end{equation}
so
\begin{equation}\label{eq:Schur_coordinate_independent}
 \frac{V_{\phi\chi}^2}{V_{\chi\chi}}
 =\frac{V_{\phi\sigma}^2}{V_{\sigma\sigma}}.
\end{equation}
For a fluctuation with frequency \(\omega\), integrating out the heavy mode gives the local resolvent
\begin{equation}\label{eq:resolvent}
 V_{\eff}''(\omega)=V_{\phi\phi}
 -V_{\phi\chi}\frac{1}{m_\chi^2-\omega^2}V_{\chi\phi}+\cdots.
\end{equation}
The derivative expansion is governed by \(\omega^2/m_\chi^2\).  The degeneration locus is therefore the endpoint of the static valley and simultaneously the endpoint of the heavy-mode derivative expansion.

Far from the degeneration locus, when \(\sigma\) varies adiabatically, the inverse-quartic potential is recovered:
\begin{equation}\label{eq:V_inv_quartic}
 V_{\rm inf}(\phi)=V_0\left(1-\frac{C}{\phi^4}+\cdots\right).
\end{equation}
The slow-roll quantities are
\begin{subequations}\label{eq:slowroll_basic}
\begin{align}
 \epsilon&=\frac12\left(\frac{4C}{\phi^5}\right)^2+\cdots,\label{eq:epsilon_basic}\\
 \eta&=-\frac{20C}{\phi^6}+\cdots,\label{eq:eta_basic}\\
 N&\simeq\frac{\phi_*^6}{24C},\label{eq:N_basic}\\
 n_s&\simeq1-\frac{5}{3N},
 \qquad r=16\epsilon_*\ll1.
 \label{eq:ns_basic}
\end{align}
\end{subequations}
The exact degeneration locus supplies the analytic boundary of validity for this single-field description.

% ============================================================
\section{Residual volume dependence of the Coulomb coefficient}\label{sec:sigma_dependent}
% ============================================================

The closed formulas above treat \(a\) as an external coefficient while the heavy volume is varied.  This is exact for the plateau term and controlled when the Coulomb fraction is small.  If the Coulomb coefficient retains residual \(\sigma\)-dependence, the critical equations are modified in a definite way.

Let
\begin{equation}\label{eq:U_asigma}
 U(\sigma,\phi)=a(\phi,\sigma)\sigma^{-2}
 +\alpha\sigma^{-4}
 -\beta\sigma^{-9/2}(\ln\sigma-L).
\end{equation}
Then
\begin{subequations}\label{eq:asigma_derivs}
\begin{align}
 U_\sigma&=\sigma^{-11/2}F(\sigma,a)+a_\sigma\sigma^{-2},\label{eq:Usigma_asigma}\\
 U_{\sigma\sigma}&=-\frac{11}{2}\sigma^{-13/2}F
 +\sigma^{-11/2}F_\sigma
 +a_{\sigma\sigma}\sigma^{-2}-4a_\sigma\sigma^{-3}.
 \label{eq:Uss_asigma}
\end{align}
\end{subequations}
Therefore the exact degeneration equations are
\begin{subequations}\label{eq:deformed_fold_F}
\begin{align}
 F+\sigma^{7/2}a_\sigma&=0,\label{eq:deformed_F1}\\
 F_\sigma+\sigma^{7/2}a_{\sigma\sigma}
 +\frac32\sigma^{5/2}a_\sigma&=0.
 \label{eq:deformed_F2}
\end{align}
\end{subequations}

A useful specialization is
\begin{equation}\label{eq:a_Ab}
 a(\phi,\sigma)=A(\phi)b(\sigma),
\end{equation}
where \(A\) contains the inflaton dependence and \(b\) contains the volume response.  With \(P\) defined in \Eq{eq:wP_main}, the two equations are
\begin{subequations}\label{eq:deformed_Ac_eqs}
\begin{align}
 U_\sigma&=A(b'\sigma^{-2}-2b\sigma^{-3})+P'=0,\label{eq:deformed_A1}\\
 U_{\sigma\sigma}&=A(b''\sigma^{-2}-4b'\sigma^{-3}+6b\sigma^{-4})+P''=0.
 \label{eq:deformed_A2}
\end{align}
\end{subequations}
Eliminating \(A\) gives
\begin{align}\label{eq:deformed_position}
0={}&P''(b'\sigma^{-2}-2b\sigma^{-3})
\nn\\
&-P'(b''\sigma^{-2}-4b'\sigma^{-3}+6b\sigma^{-4}),
\end{align}
and the critical value of the external amplitude is
\begin{equation}\label{eq:deformed_Acrit}
 A_c(\sigma)=-\frac{P'}{b'\sigma^{-2}-2b\sigma^{-3}}.
\end{equation}
For \(b=1\), \Eqs{eq:deformed_position}{eq:deformed_Acrit} reduce to \Eqs{eq:sigcrit}{eq:acrit_sigma}.

For the physical Coulomb form,
\begin{equation}\label{eq:coul_frac}
 a(\phi,\sigma)=a_0\left[1-\varepsilon_C(\phi,\sigma)\right],
 \qquad
 \varepsilon_C=\frac{\lambda(\sigma)}{\phi^4}.
\end{equation}
If \(\lambda\propto\sigma^s\) at fixed \(\phi\), then
\begin{subequations}\label{eq:sigma_dep_estimate}
\begin{align}
 \frac{|a_\sigma|\sigma}{a}
 &\simeq\frac{|s|\varepsilon_C}{1-\varepsilon_C},\label{eq:as_estimate}\\
 \frac{|a_{\sigma\sigma}|\sigma^2}{a}
 &\simeq\frac{|s(s-1)|\varepsilon_C}{1-\varepsilon_C}.
 \label{eq:ass_estimate}
\end{align}
\end{subequations}
Thus the plateau formulas are quantitatively reliable in the region
\begin{equation}\label{eq:plateau_control}
 |a_\sigma|\sigma\ll a,
 \qquad
 |a_{\sigma\sigma}|\sigma^2\ll a.
\end{equation}
In the inverse-quartic slow-roll window, write
\begin{equation}\label{eq:C_R_lambda}
 C=\mathcal R\lambda,
 \qquad
 \mathcal R=\frac{3|\Wzero|^2a_0\sigma^{-2}}{V_0}.
\end{equation}
Using \(\phi_*^6\simeq24CN\), the Coulomb fraction at horizon exit is
\begin{equation}\label{eq:epsilonC_star}
 \varepsilon_{C,*}=\frac{\lambda}{\phi_*^4}
 =\frac{C^{1/3}}{\mathcal R(24N)^{2/3}}.
\end{equation}
When this is small, the closed branch \Eq{eq:uniform_branch} governs the K\"ahler valley.  When it is not small, the exact equations \eqref{eq:deformed_F1} and \eqref{eq:deformed_F2} determine the degeneration locus.

The deformed inflaton--volume mixing is
\begin{equation}\label{eq:mixed_deformed}
 V_{\phi\sigma}=3|\Wzero|^2
 \left(a_{\phi\sigma}\sigma^{-2}-2a_\phi\sigma^{-3}\right).
\end{equation}
Cancellation of the Schur singularity requires
\begin{equation}\label{eq:mixing_cancel_condition}
 a_{\phi\sigma}\sigma-2a_\phi=\order(V_{\sigma\sigma}^{1/2})
\end{equation}
near the degeneration locus.  Without this special alignment the same fold-controlled loss of the adiabatic single-field description persists.

% ============================================================
\section{Wilsonian displacement of the degeneration locus}\label{sec:wilsonian}
% ============================================================

Near a fold, a correction that is small compared with the total potential can be decisive because both the curvature and the barrier height vanish.  Let a smooth omitted term in \(U\) be \(\Delta U\).  The shifted degeneration point obeys
\begin{equation}\label{eq:shifted_eqs}
 U_\sigma+\Delta U_\sigma=0,
 \qquad
 U_{\sigma\sigma}+\Delta U_{\sigma\sigma}=0.
\end{equation}
Linearizing around the uncorrected point and using \(U_{\sigma\sigma,c}=0\), \(U_{\sigma a,c}=-2\sigma_c^{-3}\), gives
\begin{equation}\label{eq:delta_ac}
 \delta a_c=\frac12\sigma_c^3\Delta U_{\sigma,c}.
\end{equation}
The shift in the position is
\begin{equation}\label{eq:delta_sigma}
 \delta\sigma_c=-\frac{\Delta U_{\sigma\sigma,c}+3\sigma_c^{-1}\Delta U_{\sigma,c}}
 {U_{\sigma\sigma\sigma,c}}.
\end{equation}
Equations \eqref{eq:delta_ac} and \eqref{eq:delta_sigma} are local Wilsonian response formulas.

For a power-logarithmic correction
\begin{equation}\label{eq:Delta_powerlog}
 \Delta U=A\sigma^{-p}(\ln\sigma)^q,
\end{equation}
one has
\begin{align}\label{eq:Delta_derivs}
 \Delta U_\sigma
={}&A\sigma^{-p-1}(\ln\sigma)^{q-1}
 \left[-p\ln\sigma+q\right],\nn\\
 \Delta U_{\sigma\sigma}
={}&A\sigma^{-p-2}(\ln\sigma)^{q-2}
 \big[p(p+1)(\ln\sigma)^2
 \nn\\
&-q(2p+1)\ln\sigma+q(q-1)\big].
\end{align}
Thus
\begin{equation}\label{eq:delta_ac_powerlog}
 \delta a_c=\frac{A}{2}\sigma_c^{2-p}(\ln\sigma_c)^{q-1}
 \left[-p\ln\sigma_c+q\right].
\end{equation}
The fractional displacement is
\begin{equation}\label{eq:fractional_shift}
 \frac{\delta a_c}{a_c}
 =\frac{\sigma_c^3\Delta U_{\sigma,c}}{2a_c(y)}.
\end{equation}
Because \(a_c(y)\propto(1-y)\), a smooth correction with nonzero first derivative at the endpoint is magnified as \(y\to1\).

The curvature hierarchy required for local control is
\begin{equation}\label{eq:mass_control}
 |\partial_\chi^2\Delta V|
 \ll6|\Wzero|^2\sqrt{\beta\Xi}\,\sigma_c^{-13/4}\mu^{1/2},
\end{equation}
and the barrier hierarchy is
\begin{equation}\label{eq:barrier_control}
 |\Delta V(\chi_+)-\Delta V(\chi_-)|
 \ll\frac{32}{3\sqrt{\beta\Xi}}|\Wzero|^2\sigma_c^{-3/4}\mu^{3/2}.
\end{equation}
Expanding the correction between the two extrema gives
\begin{align}\label{eq:Delta_between_extrema}
 \Delta V(\chi_+)-\Delta V(\chi_-)
 ={}&2\Delta V_\chi|_c\Delta\chi_{1/2}\nn\\
 &+\frac13\Delta V_{\chi\chi\chi}|_c\Delta\chi_{1/2}^3+\cdots .
\end{align}
where
\begin{equation}\label{eq:Dchi_half}
 \Delta\chi_{1/2}=\frac{2\sqrt6}{3\sqrt{\beta\Xi}}\sigma_c^{5/4}\mu^{1/2}.
\end{equation}
Therefore the endpoint region is controlled by local derivative inequalities, not by ordinary large-volume power counting alone.

% ============================================================
\section{Embedding into compact flux data}\label{sec:compact_flux}
% ============================================================

The continuous parameters of the two-field potential are not independent in a compact string construction.  They are functions of periods, integral fluxes, tadpole constraints, the conifold hierarchy and the K\"ahler reduction.  The local result derived above therefore becomes a set of necessary arithmetic conditions on compact type-IIB/F-theory data.  These conditions are not sufficient for a compact vacuum: sufficiency requires the full moduli spectrum, the anti-D3 effective description, and all Wilsonian corrections to remain controlled in the same region of field space.

Let \(Y_4\) be an elliptically fibered Calabi--Yau fourfold with holomorphic four-form \(\Omega_4\), integral four-form flux lattice \(H^4(Y_4,\mathbb Z)\), intersection pairing \(\eta_{IJ}\), and D3 tadpole
\begin{equation}\label{eq:Ftheory_tadpole}
 L_{\rm max}=\frac{\chi(Y_4)}{24}.
\end{equation}
For a flux vector \(N^I\in H^4(Y_4,\mathbb Z)\),
\begin{equation}\label{eq:Ftheory_W}
 W_{\rm flux}(z)=N^I\Pi_I(z),
 \qquad
 \Pi_I(z)=\int_{\Gamma_I}\Omega_4(z),
\end{equation}
and the complex-structure vacuum equations are
\begin{equation}\label{eq:Fterm_compact}
 D_AW_{\rm flux}=0.
\end{equation}
The flux contribution to the tadpole obeys
\begin{equation}\label{eq:Ftheory_tadpole_flux}
 \frac12N^I\eta_{IJ}N^J+N_{\Dthree}=L_{\rm max}.
\end{equation}
In the type-IIB orientifold limit this reduces to
\begin{equation}\label{eq:IIB_Nflux}
 N_{\rm flux}=\frac{1}{(2\pi)^4\as^2}\int_XF_3\wedge H_3,
\end{equation}
with a conifold-throat contribution \(Q_{\rm th}=KM\) and a remaining bulk contribution \cite{Bousso:2000xa,Douglas:2003um,Ashok:2003gk,Denef:2004dm,Denef:2005mm,Douglas:2006es,Saltman:2004sn,Blumenhagen:2006ci,Grimm:2024exactFlux,McAllister:2024dS,Schachner:2025overview}.

Near a conifold point, the periods of the vanishing cycle and its dual take the local form
\begin{align}\label{eq:conifold_periods}
 \Pi_A(z)&=c_Az+\order(z^2),\nn\\
 \Pi_B(z)&=\frac{c_Az}{2\pi\ii}\ln z+c_B+c_Cz+\order(z^2),
\end{align}
where \(c_A,c_B,c_C\) are determined by the regular periods of the compact Calabi--Yau.  The local fluxes in \Eq{eq:flux_ints} then give
\begin{equation}\label{eq:z_rho_compact}
 |z|=\mathcal A_z\exp\left[-\frac{2\pi K}{\gs M}\right],
 \qquad
 \rho=\frac{2\pi K}{3\gs M},
\end{equation}
where \(\mathcal A_z\) is an order-one period-dependent prefactor in the regime where the logarithmic period dominates.  The same integers that enter the tadpole therefore determine the redshift appearing in \(a_0\).

The compact dictionary is the map
\begin{align}\label{eq:compact_map}
 N^I\longmapsto
 \big(&\Wzero(N),\gs(N),M(N),K(N),\nn\\
 &\alpha(N),\beta(N),L(N)\big).
\end{align}
with
\begin{equation}\label{eq:W0_gs_compact}
 \Wzero=W_{\rm flux}\big|_{D_AW=0},
 \qquad
 \gs=\frac{1}{\Imm\tau_{\rm IIB}}.
\end{equation}
The topological contribution to the BBHL coefficient is
\begin{equation}\label{eq:xi_topological}
 \xi=-\frac{\chi(X)\zeta(3)}{2(2\pi)^3}.
\end{equation}
The remaining data in \(\alpha,\beta,L\) come from the K\"ahler reduction: string loops, logarithmic field redefinitions, higher-derivative curvature terms, orientifold data and intersection numbers.  For a fixed compactification these coefficients fix \(\theta\), and hence fix \(y\) through \(\theta=y\exp[4(1-y)/5]\).  Conversely, in a scan over compact flux data, \(y\) is a convenient coordinate on the image of the critical locus.

The control inputs used below are defined as follows.  The effective throat tadpole budget is
\begin{equation}\label{eq:Leff_def}
 L_{\eff}=L_{\rm max}-N_{\rm flux}^{\rm bulk}-N_{\Dthree}^{\rm loc},
\end{equation}
where \(N_{\rm flux}^{\rm bulk}\) denotes all flux-induced D3 charge not assigned to the conifold throat and \(N_{\Dthree}^{\rm loc}\) denotes localized D3 charge already committed to branes, curvature terms and other local sources in the chosen model.  Thus \(KM\le L_{\eff}\).  The conifold supergravity regime requires \(\gs M\gg1\).  The stronger condition
\begin{equation}\label{eq:Ccf_def}
 \gs M^2\ge C_{\rm cf}
\end{equation}
encodes the compactification-dependent lower bound required for the conifold modulus to lie inside the logarithmic region with controlled warping; \(C_{\rm cf}\) is fixed once a quantitative conifold-control prescription is chosen.  Finally, the throat must fit inside the bulk.  With \(Q_{\rm th}=KM\) and \(\Vcal_\star=\sigma_\star^{3/2}\), this is written as
\begin{equation}\label{eq:fit_condition}
 \Vcal_\star\ge C_{\rm fit}Q_{\rm th},
\end{equation}
where \(C_{\rm fit}>1\) is determined by the embedding geometry and by the convention used to define the bulk--throat transition.  Equations \eqref{eq:Ccf_def} and \eqref{eq:fit_condition} are not consequences of the local fold calculation; they are independent compact-control requirements that must be imposed before a flux vector can be considered viable.

A compact flux vector is locally admissible at a point \(0<y<1\) only if the following necessary conditions hold:
\begin{subequations}\label{eq:compact_system}
\begin{align}
 &D_AW_{\rm flux}=0,
 \label{eq:compact_vacuum}\\
 &\frac12N^I\eta_{IJ}N^J+N_{\Dthree}\le L_{\rm max},
 \label{eq:compact_tadpole}\\
 &\theta(N)=y\ee^{4(1-y)/5},
 \qquad
 \alpha(N)=\frac{9\beta(N)}{4}y\sigma_c(y)^{-1/2},
 \label{eq:compact_theta_match}\\
 &a_0(N)<a_c(y),
 \qquad
 a_0(N)=\frac{\ee^{-4\rho(N)}}{12\pi|\Wzero(N)|^2},
 \label{eq:compact_tension}\\
 &\gs M\gg1,\qquad
 \gs M^2\ge C_{\rm cf},
 \label{eq:compact_conifold_a}\\
 &\Vcal_\star\ge C_{\rm fit}KM,
 \label{eq:compact_conifold_b}\\
 &m_{3/2}<M_{s,w},
 \label{eq:compact_hierarchy_a}\\
 &\varepsilon_{C,*}\ll1
 \quad\hbox{or}\quad
 \Eqs{eq:deformed_F1}{eq:deformed_F2},
 \label{eq:compact_hierarchy}\\
 &\mu_* > \mu_{\Schur},
 \label{eq:compact_schur}\\
 &\Aobs=\frac{V_0(24N_*)^{5/3}}{192\pi^2C^{1/3}}.
 \label{eq:compact_cmb}
\end{align}
\end{subequations}
Here \(M_{s,w}\simeq\ee^{-\rho}M_s\) is the warped string scale, and \(m_{3/2}\simeq |W_0|/\Vcal\) in Planck units at leading order.  The anti-D3 effective description and possible singularity-resolution issues must also satisfy the known constraints on antibranes in warped throats \cite{Kachru:2002gs,Bena:2009xk,Bena:2010ze,Bena:2012vz,Blaback:2012nf,Danielsson:2014yga,Michel:2014lva,CohenMaldonado:2015ssa,Polchinski:2015bea,Moritz:2017xto,Gautason:2018gln,Junghans:2018gdb,Sethi:2017phn,Kachru:2018aqn}.

The critical-tension inequality is equivalently a warp inequality.  From \Eq{eq:rho_crit_y},
\begin{equation}\label{eq:compact_rho_condition}
 \frac{2\pi K(N)}{3\gs(N)M(N)}>\rho_{\min}(y).
\end{equation}
Combining this with \(KM\le L_{\eff}\) gives the integer window
\begin{equation}\label{eq:integer_window}
 K\ge\left\lceil\frac{3\gs M}{2\pi}\rho_{\min}(y)\right\rceil,
 \qquad
 K\le\left\lfloor\frac{L_{\eff}}{M}\right\rfloor .
\end{equation}
A necessary condition for any integer \(K\) to exist is therefore
\begin{equation}\label{eq:M_inequality}
 \frac{3\gs M^2}{2\pi}\rho_{\min}(y)\le L_{\eff}.
\end{equation}
If the conifold control condition \(\gs M^2\ge C_{\rm cf}\) is imposed, the local critical-tension condition is compatible with the tadpole only if
\begin{equation}\label{eq:tadpole_conifold_bound}
 \rho_{\min}(y)\le\frac{2\pi}{3}\frac{L_{\eff}}{C_{\rm cf}}.
\end{equation}
This is the local analytic form of the throat-charge obstruction.

For fixed \((\rho,L,\beta,|W_0|)\), the maximal allowed value of \(y\) follows from \(a_0=a_c(y)\).  Since
\begin{equation}\label{eq:ac_constantC}
 a_c(y)=C_a\ee^{-4y}(1-y),
 \qquad
 C_a=\frac{9\beta}{10}\ee^{-5L/2-14/9},
\end{equation}
set
\begin{equation}\label{eq:Z_def}
 Z=\frac{4\ee^4a_0}{C_a}.
\end{equation}
Then
\begin{equation}\label{eq:ymax}
 y_{\rm max}=1-\frac14\LambertW_0(Z),
\end{equation}
provided \(0<y_{\rm max}<1\).  Since \(a_0\propto |W_0|^{-2}\),
\begin{equation}\label{eq:W0_sensitivity_correct}
 \frac{\partial y_{\rm max}}{\partial\ln|W_0|}
 =\frac12\frac{\LambertW_0(Z)}{1+\LambertW_0(Z)}>0.
\end{equation}
Increasing \(|W_0|\) decreases the plateau anti-D3 tension parameter and enlarges the admissible interval in \(y\), with all other quantities held fixed.

The one-volume reduction must be embedded in the full K\"ahler mass matrix.  Let \(\varphi^I=(\sigma,\psi^a)\), where \(\psi^a\) are orthogonal K\"ahler-sector fields, and let \(E_A{}^I\) be a vielbein for the K\"ahler field-space metric, \(E_A{}^IE_B{}^JG_{IJ}=\delta_{AB}\).  The physical Hessian in the orthonormal frame is
\begin{equation}\label{eq:physical_Hessian_block}
 {\cal H}_{AB}=E_A{}^IE_B{}^J\nabla_I\nabla_JV.
\end{equation}
Writing \(A=(\chi,a)\), the K\"ahler block is
\begin{equation}\label{eq:Hessian_block}
 {\cal H}_{AB}=\begin{pmatrix}
 {\cal H}_{\chi\chi} & {\cal H}_{\chi b}\\
 {\cal H}_{a\chi} & {\cal H}_{ab}
 \end{pmatrix}.
\end{equation}
If the orthogonal block is invertible, the physical eigenvalue controlling the volume direction after the orthogonal K\"ahler fields are integrated out is
\begin{equation}\label{eq:nested_schur_main}
 {\cal H}_{\chi\chi}^{\rm red}
 ={\cal H}_{\chi\chi}-{\cal H}_{\chi a}({\cal H}^{-1})^{ab}{\cal H}_{b\chi}.
\end{equation}
The one-volume degeneration derived above is the first compact instability only if
\begin{equation}\label{eq:orthogonal_condition}
 m_{\perp}^2\gg m_\chi^2,
 \qquad
 |{\cal H}_{\chi a}({\cal H}^{-1})^{ab}{\cal H}_{b\chi}|
 \ll |{\cal H}_{\chi\chi}|
\end{equation}
in the region of interest.  Otherwise the true compact degeneration locus is the zero of the reduced physical Hessian \eqref{eq:nested_schur_main}, not the one-volume expression \eqref{eq:fold_intro}.

% ============================================================
\section{Flux arithmetic, hierarchies and scalar normalization}\label{sec:global_constraints}
% ============================================================

The compact constraints may be organized as a sequence of necessary filters.  First, the flux vector must solve \(D_AW=0\) and satisfy the D3 tadpole.  Second, the conifold fluxes must generate a hierarchy large enough to satisfy \(a_0<a_c(y)\) while keeping the throat in the supergravity regime.  Third, the physical K\"ahler mass matrix must have the one-volume eigenvalue as the first eigenvalue approaching zero.  Fourth, the scalar normalization and the Schur bound must be compatible with the same point in parameter space.

The scalar amplitude in the inverse-quartic regime follows directly from \Eq{eq:V_inv_quartic}.  This derivation assumes \(C/\phi_*^4\ll1\), \(V_*\simeq V_0\), \(\phi_*\gg\phi_{\rm end}\), and \(M_{\rm P}=1\).  Using \(\Aobs=V/(24\pi^2\epsilon)\) and \(\phi_*^6=24CN_*\), one finds
\begin{equation}\label{eq:As_formula}
 \Aobs=\frac{V_0(24N_*)^{5/3}}{192\pi^2C^{1/3}}.
\end{equation}
Equivalently,
\begin{equation}\label{eq:C_from_As}
 C=\left[\frac{V_0(24N_*)^{5/3}}{192\pi^2\Aobs}\right]^3.
\end{equation}
In the plateau approximation,
\begin{equation}\label{eq:V0_plateau}
 V_0\simeq3|\Wzero|^2a_0\sigma_\star^{-2}+V_{\rm pert}(\sigma_\star),
\end{equation}
where \(\sigma_\star\) is the stable branch solution at horizon exit.  The Coulomb coefficient \(C\) is determined by \(\lambda\), \(a_0\), \(W_0\), and the volume response through \Eq{eq:C_R_lambda}.  The scalar amplitude therefore constrains the same compact data that enter the critical-tension condition.

Let
\begin{equation}\label{eq:mu_star}
 \mu_*=a_c-a(\phi_*).
\end{equation}
The local K\"ahler valley is under adiabatic control only if
\begin{equation}\label{eq:mu_star_schur}
 \mu_*\gg\mu_{\Schur},
\end{equation}
where \(\mu_{\Schur}\) is given by \Eq{eq:mu_schur_general} or \Eq{eq:mu_schur_intrinsic}.  In addition, the physical K\"ahler mass must lie between the Hubble scale and the compactification cutoff in the controlled single-field regime,
\begin{equation}\label{eq:mass_hierarchy}
 H_*^2\ll m_\chi^2\ll M_{\rm KK}^2,
\end{equation}
with \(m_\chi^2\) evaluated by \Eq{eq:mchi_uniform}.  Near the degeneration locus this hierarchy fails from the left because \(m_\chi^2\propto\mu^{1/2}\).

The critical tension parameter also determines the local decay barrier.  The semiclassical tunneling rate cannot be inferred from \(\Delta V_B\) alone, but the vanishing of \Eq{eq:DV_uniform} is a necessary signal that metastability disappears.  Any Euclidean-bounce analysis near the endpoint must use the cubic normal form \Eq{eq:canonical_normal_form} and the canonical separation \Eq{eq:Dchi_uniform}, not merely the value of the potential at the minimum \cite{Coleman:1977py,Hawking:1981fz}.

Putting the ingredients together, a compact flux vector can be retained by the analytic local filter only if there exists a real \(y\in(0,1)\) such that
\begin{widetext}
\begin{equation}\label{eq:final_filter_set}
\begin{aligned}
 \theta(N)&=y\ee^{4(1-y)/5},
&
 \alpha(N)&=\frac{9\beta(N)}{4}y\sigma_c(y)^{-1/2},
\\
 a_0(N)&<a_c(y),
&
 \frac{3\gs(N)M(N)^2}{2\pi}\rho_{\min}(y)&\le L_{\eff}(N),
\\
 \gs M&\gg1,
&
 \gs M^2&\ge C_{\rm cf},
\\
 \Vcal_\star&\ge C_{\rm fit}KM,
&
 \varepsilon_{C,*}&\ll1\quad\hbox{or the deformed equations hold},
\\
 \mu_*&\gg\mu_{\Schur},
&
 H_*^2&\ll m_\chi^2\ll M_{\rm KK}^2,
\\
 {\cal H}_{\chi\chi}^{\rm red}&>0,
&
 0&<y<1 .
\end{aligned}
\end{equation}
\end{widetext}
The last inequality is required throughout the trajectory until the one-volume degeneration is reached.  This is the exact necessary local condition implied by the leading perturbative brane--antibrane potential, supplemented by compact-flux, conifold, hierarchy and K\"ahler-mass requirements.  It is not a proof of compact metastability; it is the analytic filter that any numerical compact search must satisfy before a candidate can be promoted to a full compact solution.

\section{Numerical realization of the critical locus}
\label{sec:numerical_realization}

We now evaluate the analytic formulae on a concrete benchmark of the reduced effective theory.  The purpose of this section is not to construct a compact vacuum, but to verify in one numerical realization the exact critical locus of \Sec{sec:critical_locus}, the elementary positive-tension uniformization of \Sec{sec:uniformization}, the saddle-node normal form of \Sec{sec:normal_form}, the Schur-complement singularity of \Sec{sec:schur}, and the compact flux filtration of \Secs{sec:compact_flux}{sec:global_constraints}.  All dimensionful quantities are expressed in reduced Planck units, \(M_{\rm P}=1\).  The plotted variables \(\sigma/\sigma_c\), \(a/a_c\), \(\mu/a_c\), \(K/M\), and \(\rho\) are dimensionless, while \(M\) and \(K\) are integer flux quanta.

The benchmark input is displayed in \Tab{tab:numerical_input}.  The value of \(\theta\) is fixed first, and the branch coordinate \(y_c\) is obtained from \Eq{eq:theta_y}.  The coefficient \(\alpha\) is then determined by \Eq{eq:alpha_y}, so that the benchmark lies exactly on the analytic critical branch.  This procedure is the numerical counterpart of the compact data map discussed in \Sec{sec:compact_flux}: after the compact reduction has produced \((\alpha,\beta,L)\), the branch coordinate is not an additional modulus but the solution of the critical-branch equation.

\begin{table}[t]
\centering
\caption{
Benchmark used in the numerical realization.  The parameters \(g_s,L,\beta,\theta,|W_0|\) determine the local reduced potential \Eq{eq:U_def_intro}.  The quantities \(L_{\rm eff},C_{\rm cf}\) and \(C_{\rm fit}\) define the necessary throat-flux filters used in \Sec{sec:global_constraints}.
}
\label{tab:numerical_input}
\begin{ruledtabular}
\begin{tabular}{lc}
Quantity & Value \\
\hline
\(g_s\) & \(0.2\) \\
\(L\) & \(5.0\) \\
\(\beta\) & \(1.0\) \\
\(\theta\) & \(0.55\) \\
\(|W_0|\) & \(10^{-4}\) \\
\(L_{\rm eff}\) & \(320\) \\
\(C_{\rm cf}\) & \(46.0\) \\
\(C_{\rm fit}\) & \(1.1\) \\
\(M_{\rm max}=K_{\rm max}\) & \(80\) \\
\(\phi_c\) & \(1.0\) \\
\end{tabular}
\end{ruledtabular}
\end{table}

Solving \Eq{eq:theta_y} gives
\begin{equation}
 y_c=0.3189688111,
 \qquad
 \Xi_c=5-4y_c=3.724124756 .
\end{equation}
Using \Eqs{eq:sigma_y}{eq:ac_y}, the corresponding critical point is
\begin{equation}
 \sigma_c=460.6194988,
 \qquad
 a_c=1.3460268\times10^{-7}.
\end{equation}
The perturbative coefficient determined by \Eq{eq:alpha_y} is
\begin{equation}
 \alpha=0.03343948486 .
\end{equation}
The local cubic coefficient has the sign required by \Eq{eq:Fss_negative},
\begin{equation}
 F_{\sigma\sigma,c}
 =-3.94931798\times10^{-5}<0,
\end{equation}
so the lower-\(\sigma\) stationary point is the K\"ahler minimum and the higher-\(\sigma\) stationary point is the decompactification barrier.

\Tab{tab:numerical_audit} gives the numerical residuals of the analytic construction.  The direct solution of \(F=F_\sigma=0\), the closed formulae \Eqs{eq:acrit_sigma}{eq:sigcrit}, the uniformized branch \Eqs{eq:theta_y}{eq:ac_y}, and the warped-hierarchy expression \Eq{eq:rho_c_y} agree to machine precision.  This confirms that the numerical plots below are not produced by fitting the stationary branches; they are evaluations of the exact formulae derived in the preceding sections.

\begin{table*}[t]
\centering
\caption{
Audit of the defining equations and closed formulae at the numerical critical point.  The residuals are dimensionless.  The agreement between the direct critical equations, the closed expressions, and the uniformized \(y\)-branch verifies the algebraic solution used in the figures.
}
\label{tab:numerical_audit}
\begin{ruledtabular}
\begin{tabular}{lcc}
Test & Equation or identity & Residual \\
\hline
Stationarity & \(F(\sigma_c,a_c)=0\) & \(1.9984\times10^{-15}\) \\
Vanishing curvature & \(F_\sigma(\sigma_c,a_c)=0\) & \(1.7347\times10^{-18}\) \\
Critical equation & \Eq{eq:sigcrit} & \(4.4409\times10^{-16}\) \\
Critical tension & \Eq{eq:acrit_sigma} & \(0\) \\
Lambert critical equation & \Eq{eq:Lambert_u} & \(4.4409\times10^{-16}\) \\
Cubic coefficient identity & \Eq{eq:Fss_uniform} & \(6.7763\times10^{-21}\) \\
Uniformized \(\theta\) & \Eq{eq:theta_y} & \(0\) \\
Uniformized \(\sigma_c\) & \Eq{eq:sigma_y} & \(3.4106\times10^{-13}\) \\
Uniformized \(\alpha\) & \Eq{eq:alpha_y} & \(0\) \\
Uniformized \(a_c\) & \Eq{eq:ac_y} & \(1.3235\times10^{-22}\) \\
Warp exponent & \Eq{eq:rho_c_y} & \(0\) \\
\end{tabular}
\end{ruledtabular}
\end{table*}

\subsection{Local geometry of the K\"ahler-valley endpoint}
\label{subsec:numerical_local_geometry}

\begin{figure*}
\centering
\includegraphics[width=0.94\textwidth]{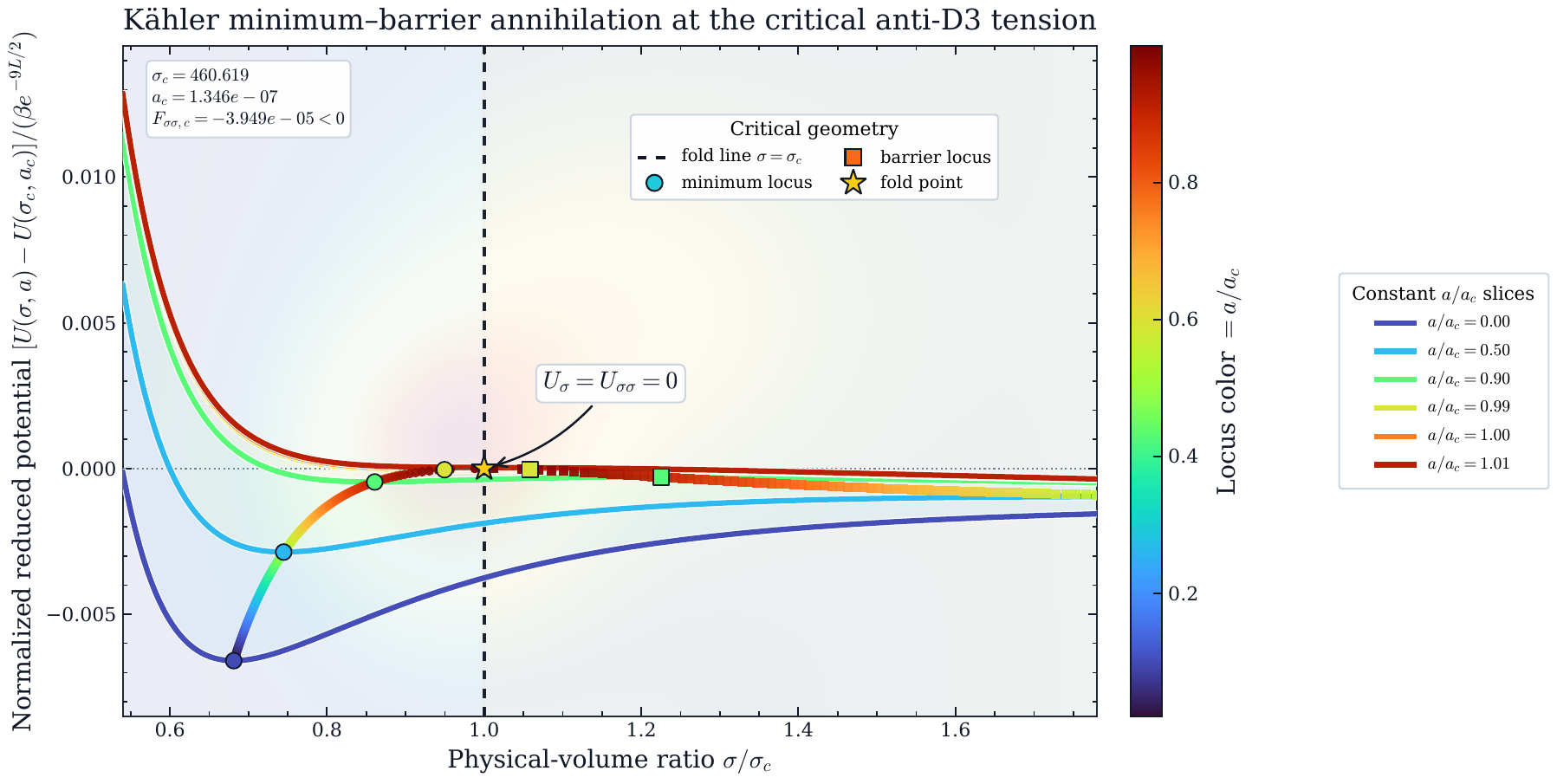}
\caption{
K\"ahler minimum--barrier annihilation at the critical anti-D3 tension.  The plotted quantity is the normalized reduced potential
\([U(\sigma,a)-U(\sigma_c,a_c)]/(\beta e^{-9L/2})\).  For \(a<a_c\), the lower-\(\sigma\) stationary point is the K\"ahler minimum and the higher-\(\sigma\) stationary point is the decompactification barrier.  At \(a=a_c\), the two branches meet at \(U_\sigma=U_{\sigma\sigma}=0\).  For \(a>a_c\), the metastable physical-volume trajectory is absent.
}
\label{fig:numerical_volume_slices}
\end{figure*}

\Fig{fig:numerical_volume_slices} shows one-dimensional sections of the reduced potential \Eq{eq:U_def_intro} near the critical point.  The plotted potential is normalized by \(\beta e^{-9L/2}\), and the horizontal coordinate is the physical-volume ratio \(\sigma/\sigma_c\).  For \(a<a_c\), the stationary condition \(U_\sigma=0\) has two nearby roots.  The lower-\(\sigma\) root is the local K\"ahler minimum, while the higher-\(\sigma\) root is the decompactification barrier.  As \(a/a_c\) approaches unity, the two roots coalesce at the point marked by
\[
 U_\sigma=U_{\sigma\sigma}=0 .
\]
For \(a>a_c\), the local minimum and barrier are absent.  The figure is therefore the direct numerical realization of \Eq{eq:foldconds}.

\begin{figure*}
\centering
\includegraphics[width=0.94\textwidth]{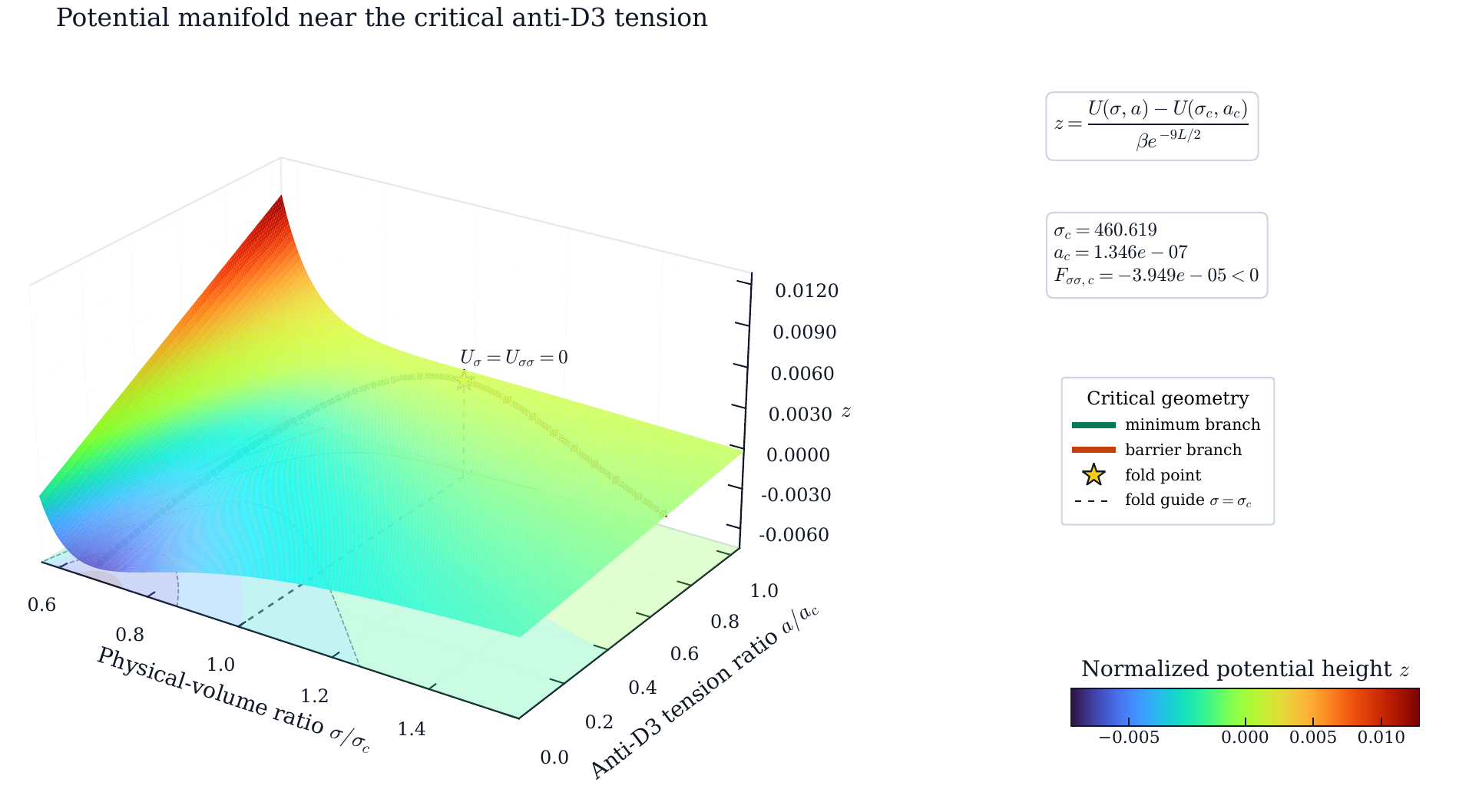}
\caption{
Potential manifold near the critical anti-D3 tension. The surface shows
\(\displaystyle
z=\frac{U(\sigma,a)-U(\sigma_c,a_c)}{\beta e^{-9L/2}}
\)
over the \((\sigma/\sigma_c,a/a_c)\) plane. The minimum and barrier branches are stationary trajectories on this surface and terminate at the fold point. The geometry gives a numerical visualization of the simultaneous equations \(U_\sigma=U_{\sigma\sigma}=0\).
}
\label{fig:numerical_potential_manifold}
\end{figure*}

The same structure is shown as a potential manifold in \Fig{fig:numerical_potential_manifold}.  The two stationary branches are embedded in the \((\sigma/\sigma_c,a/a_c)\) surface and terminate at the same critical point.  This view is useful because it makes clear that the local loss of metastability is a codimension-one degeneration in the space of anti-D3 tension parameters, not an artifact of choosing selected one-dimensional slices.  The fold guide at \(\sigma=\sigma_c\) is the numerical counterpart of the local expansion variables \(q=\sigma-\sigma_c\) and \(\mu=a_c-a\) introduced in \Eq{eq:local_vars}.

Together, it shows that the sign result \Eq{eq:Fss_negative} has the expected physical interpretation.  The stable branch lies at \(q<0\), the decompactification barrier lies at \(q>0\), and their separation collapses as \(\mu^{1/2}\) as derived in \Eq{eq:q_pm}.

\subsection{Normal-form scaling and Schur-complement singularity}
\label{subsec:numerical_normal_form}

\begin{figure*}
\centering
\includegraphics[width=0.94\textwidth]{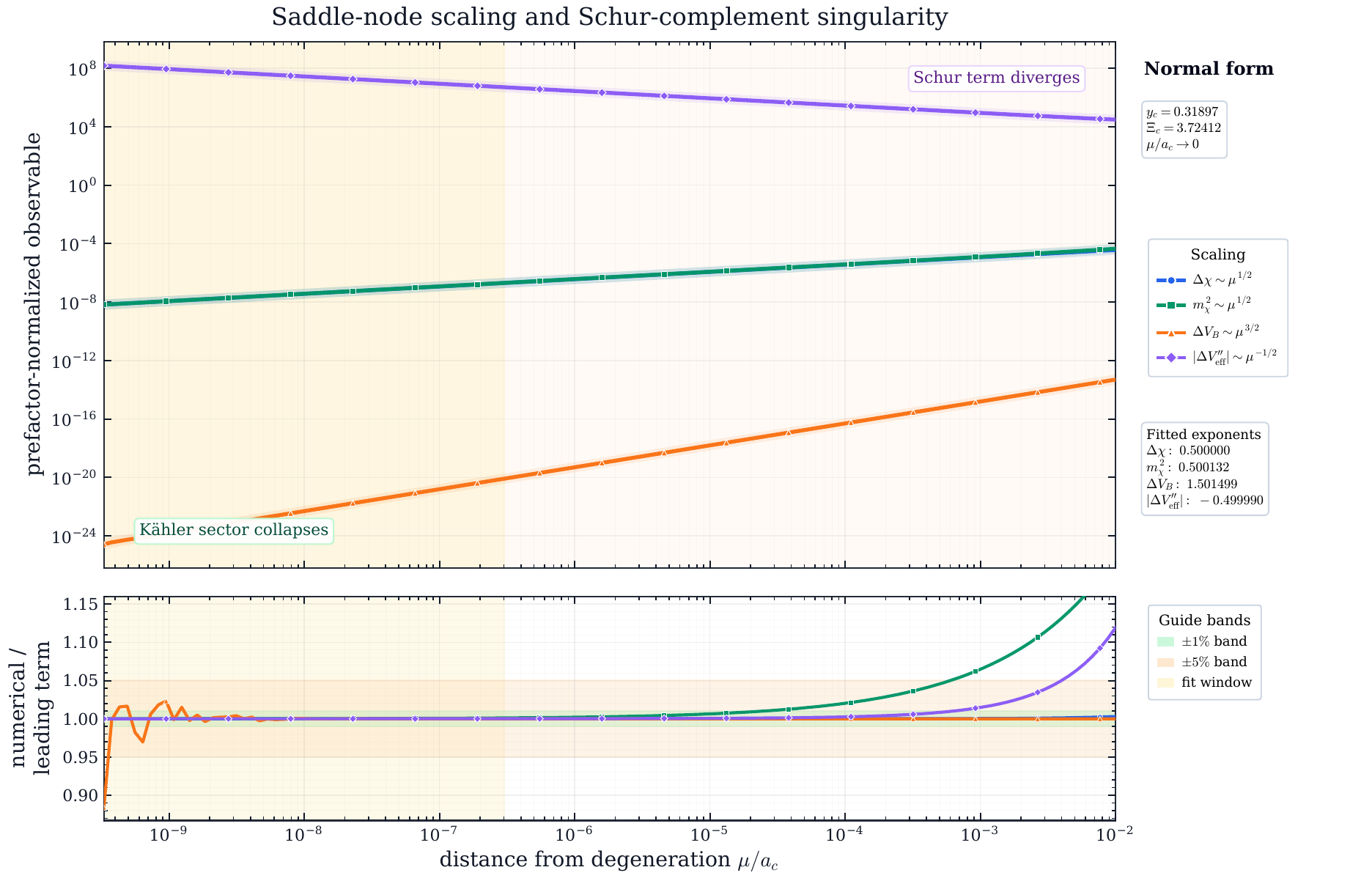}
\caption{
Saddle-node scaling and Schur-complement singularity.  The top panel shows the exact numerical observables divided by their analytic prefactors.  The plotted curves confirm
\(\Delta\chi\sim\mu^{1/2}\), \(m_\chi^2\sim\mu^{1/2}\), \(\Delta V_B\sim\mu^{3/2}\), and
\(|\Delta V_{\rm eff}''|\sim\mu^{-1/2}\).  The lower panel shows the ratio of the numerical result to the leading analytic term.  The fit window is restricted to the region in which the stationary points are close to the fold while numerical subtraction in \(\Delta V_B\) remains stable.
}
\label{fig:numerical_schur_collapse}
\end{figure*}

The next numerical test concerns the local normal form.  \Fig{fig:numerical_schur_collapse} compares the exact numerical stationary points with the leading analytic coefficients in \Eq{eq:local_coeff_uniform} and the Schur residue in \Eq{eq:Schur_residue_uniform}.  The upper panel plots the exact observables after dividing out their analytic prefactors.  The remaining dependence is the universal power of \(\mu=a_c-a\).  The lower panel divides each numerical result by the full leading asymptotic expression, so that convergence to unity verifies the coefficient as well as the exponent.

The fitted exponents are summarized in \Tab{tab:numerical_exponents}.  The first three rows test the K\"ahler-sector normal form \Eq{eq:fold_scaling}; the last row tests the Schur-complement pole \Eq{eq:Schur_residue_uniform}.  The results reproduce the analytic values with percent-level or better accuracy.  The barrier height is the most sensitive observable because it is obtained by subtracting the potential at two nearby stationary points.  Its fitted exponent nevertheless remains close to the predicted \(3/2\), and the ratio panel in \Fig{fig:numerical_schur_collapse} shows convergence to the leading expression outside the cancellation-dominated edge of the scan.

\begin{table}[t]
\centering
\caption{
Numerical exponents near the degeneration locus.  The exact predictions follow from \Eq{eq:local_coeff_uniform} and \Eq{eq:Schur_residue_uniform}.  The fitted exponents are obtained from the computed stationary branches as \(\mu/a_c\to0^+\).
}
\label{tab:numerical_exponents}
\begin{ruledtabular}
\begin{tabular}{lcc}
Observable & Prediction & Fit \\
\hline
\(\Delta\chi\) & \(1/2\) & \(0.49999985\) \\
\(m_\chi^2\) & \(1/2\) & \(0.50013155\) \\
\(\Delta V_B\) & \(3/2\) & \(1.50149930\) \\
\(|\Delta V_{\eff}''|\) & \(-1/2\) & \(-0.49998981\) \\
\end{tabular}
\end{ruledtabular}
\end{table}

The physical implication is important.  The canonical distance \(\Delta\chi\), the K\"ahler mass \(m_\chi^2\), and the barrier height \(\Delta V_B\) vanish as the degeneration point is approached, while the adiabatic single-field correction diverges.  Thus the numerical realization confirms the analytic statement of \Sec{sec:schur}: the endpoint of the heavy K\"ahler valley and the breakdown of the adiabatic single-field Hessian are controlled by the same vanishing denominator \(V_{\sigma\sigma}\).  Unless \(a_{\phi,c}\) is suppressed, the Schur term becomes large before the local potential reaches the fold.

\subsection{Integer throat-flux filtration}
\label{subsec:numerical_flux_filtration}

The local critical-tension condition becomes an arithmetic constraint after imposing the throat relation \Eq{eq:rho_def}.  For the benchmark above, \Eq{eq:rho_c_y} gives
\begin{equation}
 \rho_{\min}=7.652995675,
\end{equation}
and hence \Eq{eq:flux_ratio_bound} gives
\begin{equation}
 \frac{K}{M}>0.730807255 .
\end{equation}
The continuous minimum tadpole cost implied by the warping and conifold filters is
\begin{equation}
 Q_{\rm th}^{\rm cont}\simeq 168.086,
\end{equation}
while the integer scan finds the first admissible throat at
\begin{equation}
 (M,K)=(16,12),
 \qquad
 KM=192,
 \qquad
 \rho=7.85398 .
\end{equation}
The remaining tadpole margin at this point is
\begin{equation}
 L_{\rm eff}-KM=128 .
\end{equation}

\begin{figure*}
\centering
\includegraphics[width=0.94\textwidth]{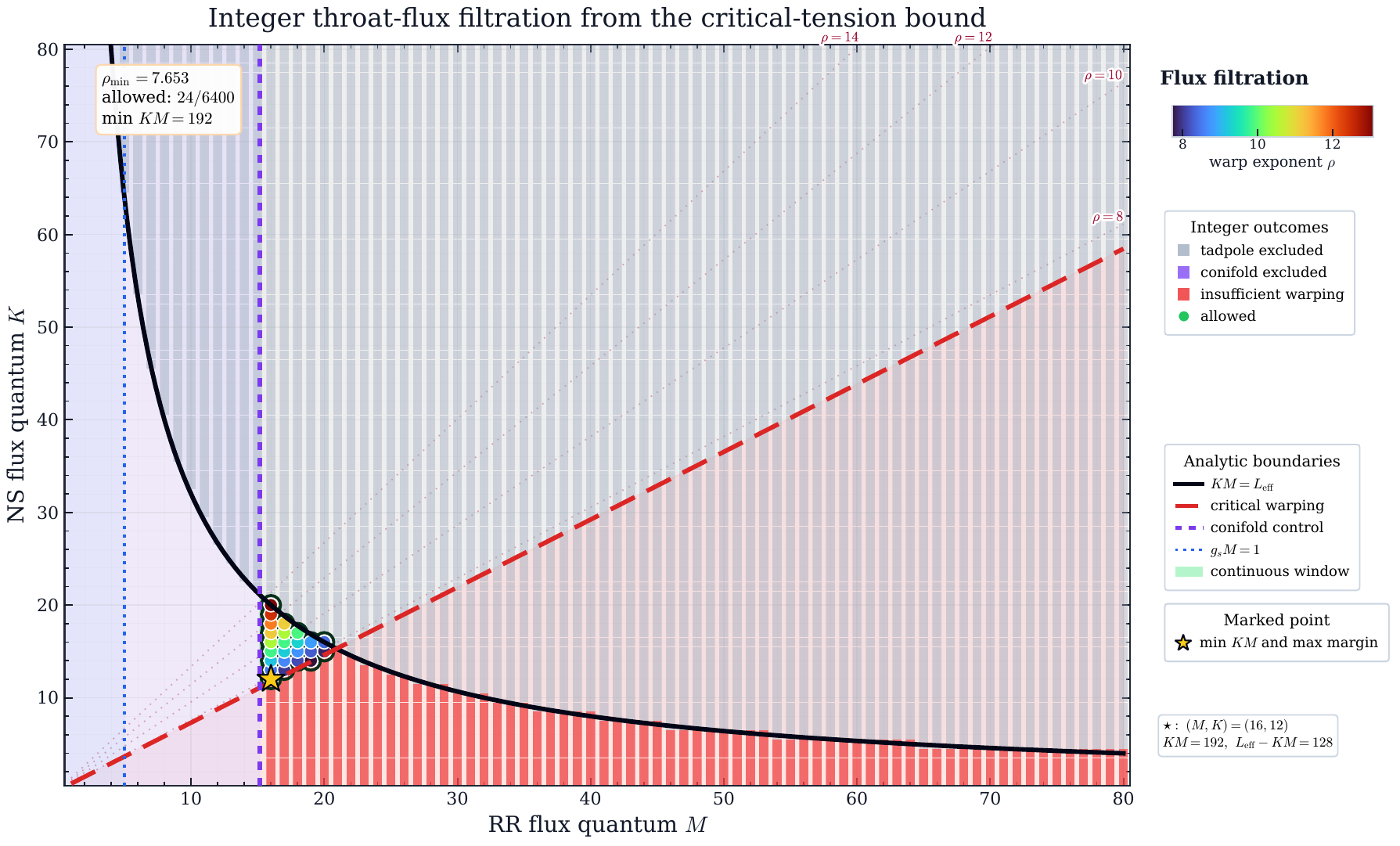}
\caption{
Integer throat-flux filtration from the critical-tension bound.  The scan covers \(1\le M,K\le80\).  The black curve is \(KM=L_{\rm eff}\), the red dashed line is the critical warped-hierarchy condition from \Eq{eq:flux_ratio_bound}, and the vertical control lines indicate the conifold and \(g_sM\) constraints.  The green region is the continuous necessary window.  The colored lattice points show the first failed condition for each integer flux pair.  Only \(24\) out of \(6400\) integer pairs pass the displayed filters.  These points are necessary-filter survivors, not compact vacua.
}
\label{fig:numerical_flux_lattice}
\end{figure*}

\Fig{fig:numerical_flux_lattice} displays the full integer scan over \(1\le M,K\le80\).  The black curve is the tadpole boundary \(KM=L_{\rm eff}\), the red dashed line is the critical-warping condition \Eq{eq:flux_ratio_bound}, and the vertical control lines represent the conifold-curvature and \(g_sM\) constraints.  The continuous green region is the necessary real-valued window, but the actual allowed set consists only of the integer lattice points inside that region which also satisfy the conifold and tadpole restrictions.

The filtration counts are given in \Tab{tab:numerical_flux_counts}.  Most points are removed by the tadpole condition.  The conifold-curvature and warping constraints then remove the low-\(M\) and low-\(K/M\) regions, respectively.  In this benchmark the bulk-volume filter does not remove any additional points after the preceding cuts.  This is not a general statement; it reflects the chosen \(C_{\rm fit}\), \(L_{\rm eff}\), and scan range.

\begin{table}[t]
\centering
\caption{
Outcome of the integer throat-flux scan over \(1\le M,K\le80\).  The categories are mutually exclusive and record the first failed condition in the implemented filter sequence.  The allowed points satisfy the displayed necessary local and throat-control conditions, but are not asserted to be compact solutions.
}
\label{tab:numerical_flux_counts}
\begin{ruledtabular}
\begin{tabular}{lc}
Flux class & Number of points \\
\hline
Tadpole excluded & \(5191\) \\
Conifold excluded & \(711\) \\
Insufficient warping & \(474\) \\
Bulk-volume excluded & \(0\) \\
Allowed by displayed filters & \(24\) \\
\hline
Total & \(6400\) \\
\end{tabular}
\end{ruledtabular}
\end{table}

The allowed set is narrow:
\begin{equation}
 16\le M\le20,
 \qquad
 12\le K\le20,
 \qquad
 192\le KM\le320.
\end{equation}
Thus the analytic critical-tension condition does not merely impose a mild real inequality.  Once combined with integer flux quantization and the tadpole budget, it produces a sparse filtration of throat data.  This is the numerical version of the necessary compact filter written in \Eq{eq:final_filter_set}.  A full compact model would still have to solve the period equations \(D_AW=0\), evaluate the perturbative K\"ahler dictionary producing \((\alpha,\beta,L)\), verify the full physical K\"ahler-sector Schur condition \Eq{eq:nested_schur_main}, and check the Wilsonian inequalities of \Sec{sec:wilsonian}.  The present scan demonstrates only that the local analytic bound has sharp arithmetic consequences for the throat-flux sector.

The numerical realization supports four linked statements.  First, the direct solution of the critical equations \(F=F_\sigma=0\) agrees with the closed analytic surface \Eqs{eq:acrit_sigma}{eq:sigcrit}.  Second, the elementary \(y\)-coordinate in \Eq{eq:uniform_branch} reproduces the same critical point without numerical root finding.  Third, the physical-volume minimum and decompactification barrier have the saddle-node scaling predicted by \Eq{eq:local_coeff_uniform}.  Fourth, the same approach to the degeneration locus produces the Schur-complement singularity \Eq{eq:Schur_residue_uniform}, showing that the adiabatic single-field description fails unless the inflaton--volume mixing is suppressed.

The compact-flux scan then translates this local result into a finite integer filtration.  For the benchmark considered here, only \(24/6400\) throat-flux pairs pass the displayed necessary conditions.  This is a numerical illustration of the logic of \Sec{sec:global_constraints}: the critical anti-D3 tension defines a calculable local boundary, and compact data must land on the safe side of that boundary while also satisfying tadpole, conifold, hierarchy and full K\"ahler-mass constraints.  The allowed flux points in \Fig{fig:numerical_flux_lattice} should therefore be interpreted as candidates for further compact analysis, not as vacua.

\clearpage
\section{Conclusion}\label{sec:conclusion}

This paper has identified and solved the local obstruction that controls the existence of the physical-volume trajectory in perturbative brane--antibrane inflation.  The starting point was the leading nilpotent effective potential
\begin{equation}
 V=3|\Wzero|^2U,
 \qquad
 U(\sigma,a)=a\sigma^{-2}+\alpha\sigma^{-4}
 -\beta\sigma^{-9/2}(\ln\sigma-L),
\end{equation}
whose anti-D3 term scales more slowly with the physical volume than the perturbative stabilizing terms.  The resulting competition implies that increasing the anti-D3 tension parameter cannot be done indefinitely: the K\"ahler minimum and the adjacent decompactification barrier meet at a real codimension-one degeneration locus.  This locus is determined exactly by the simultaneous equations
\begin{equation}
 U_\sigma(\sigma_c,a_c)=0,
 \qquad
 U_{\sigma\sigma}(\sigma_c,a_c)=0,
\end{equation}
or equivalently by \(F(\sigma_c,a_c)=F_\sigma(\sigma_c,a_c)=0\).  The local problem is therefore not a numerical minimization problem but an analytically solvable fold of the reduced scalar potential.

The first main result is the closed critical surface
\begin{align}
 a_c(\sigma_c)
 &=\frac{9\beta}{10}\sigma_c^{-5/2}
 -\frac{2\alpha}{5}\sigma_c^{-2},
 \\
 \ln\sigma_c-L
 &=\frac{28}{45}
 +\frac{32}{45}\frac{\alpha}{\beta}\sqrt{\sigma_c}.
\end{align}
The positive-tension component of this surface admits the elementary global parameterization
\begin{align}
 \theta(y)&=y\exp\!\left[\frac45(1-y)\right],
 \\
 \sigma_c(y)&=\exp\!\left[
 L+\frac{20}{9}-\frac85(1-y)
 \right],
 \\
 a_c(y)&=\frac{9\beta}{10}
 \exp\!\left[
 -\frac{5L}{2}-\frac{14}{9}-4y
 \right](1-y),
\end{align}
with
\begin{equation}
 0<y<1.
\end{equation}
This replaces the Lambert-function branch structure and numerical root finding by one real coordinate on the physical critical branch.  For a fixed effective theory, \(y\) is fixed by the microscopic data through \(\theta=y\exp[4(1-y)/5]\); it is not an additional modulus.  Across a family of flux data, however, \(y\) becomes a convenient coordinate on the image of the compact K\"ahler reduction.

The second main result is the explicit normal form of the degeneration.  On the elementary branch all local cubic coefficients are controlled by
\begin{equation}
 \Xi(y)=5-4y,
 \qquad 1<\Xi<5.
\end{equation}
Writing
\begin{equation}
 \mu=a_c-a,
\end{equation}
the metastable side is \(\mu>0\).  The canonical separation between the K\"ahler minimum and the decompactification barrier, the canonically normalized K\"ahler mass, and the barrier height obey
\begin{align}
 \Delta\chi
 &=\frac{4\sqrt6}{3\sqrt{\beta\Xi}}
 \sigma_c^{5/4}\mu^{1/2}
 +\order(\mu^{3/2}),
 \\
 m_\chi^2
 &=6|\Wzero|^2\sqrt{\beta\Xi}\,
 \sigma_c^{-13/4}\mu^{1/2}
 +\order(\mu),
 \\
 \Delta V_B
 &=\frac{32}{3\sqrt{\beta\Xi}}
 |\Wzero|^2\sigma_c^{-3/4}\mu^{3/2}
 +\order(\mu^2).
\end{align}
Thus the physical-volume trajectory disappears by a saddle-node degeneration with the expected hierarchy
\begin{equation}
 \Delta\chi\sim\mu^{1/2},
 \qquad
 m_\chi^2\sim\mu^{1/2},
 \qquad
 \Delta V_B\sim\mu^{3/2}.
\end{equation}
The coefficients are not arbitrary normal-form constants: they are fixed explicitly by the string-theory parameters \((\beta,L,\Wzero,y)\).

The third main result is the relation between the degeneration locus and the breakdown of the adiabatic single-field description.  Along the K\"ahler valley, integrating out the physical volume gives the Schur-complement correction
\begin{equation}
 \frac{\dd^2V_{\eff}}{\dd\phi^2}
 =
 V_{\phi\phi}
 -
 \frac{V_{\phi\sigma}^2}{V_{\sigma\sigma}}.
\end{equation}
Since \(V_{\sigma\sigma}\to0\) at the same locus where the K\"ahler minimum collides with the decompactification barrier, the adiabatic Hessian develops
\begin{equation}
 \Delta V_{\eff}''
 =
 -\frac{4|\Wzero|^2a_{\phi,c}^{\,2}}
 {\sqrt{\beta\Xi(y)}\,\sigma_c^{3/4}}
 \mu^{-1/2}
 +\order(\mu^0),
\end{equation}
unless the inflaton--volume mixing vanishes with sufficient strength at the critical point.  Hence the local endpoint is simultaneously a boundary for static K\"ahler metastability and for the derivative expansion underlying the single-field effective theory.

The fourth result is the compact-flux interpretation.  The critical-tension condition gives a necessary warped-hierarchy bound,
\begin{equation}
 \rho>\rho_{\min}(y),
\end{equation}
and using the throat relation
\begin{equation}
 \rho=\frac{2\pi K}{3g_sM}
\end{equation}
one obtains the integer condition
\begin{equation}
 K\ge
 \left\lceil
 \frac{3g_sM}{2\pi}\rho_{\min}(y)
 \right\rceil,
 \qquad
 K\le
 \left\lfloor
 \frac{L_{\eff}}{M}
 \right\rfloor .
\end{equation}
This expresses the local K\"ahler-valley requirement directly in terms of throat-flux data.  Together with the tadpole budget, conifold-control inequalities, scalar-amplitude normalization, and the full K\"ahler-sector Schur condition, it gives an analytic necessary filter for compact type-IIB/F-theory flux data.  These conditions are deliberately not stated as sufficient.  A compact vacuum would require the complete solution of the period equations, the K\"ahler reduction that produces \((\alpha,\beta,L)\), the full physical mass matrix, anti-D3 control, and the Wilsonian hierarchy conditions to hold simultaneously.

The numerical realization in \Sec{sec:numerical_realization} verifies the analytic structure in a concrete benchmark.  The critical equations \(F=F_\sigma=0\), the closed expressions for \(\sigma_c\) and \(a_c\), and the uniformized \(y\)-branch agree to numerical precision.  The potential slices and the two-parameter potential manifold show the annihilation of the K\"ahler minimum and the decompactification barrier.  The normal-form plot confirms the exponents \(1/2\), \(1/2\), \(3/2\), and \(-1/2\) for \(\Delta\chi\), \(m_\chi^2\), \(\Delta V_B\), and \(|\Delta V_{\eff}''|\), respectively.  Finally, the integer throat-flux scan shows that the analytic local condition becomes a sparse arithmetic filtration once the tadpole, conifold, and warped-hierarchy constraints are imposed.  In the benchmark studied there, only \(24\) of \(6400\) integer pairs in the displayed scan window survive the necessary filters.

Several limitations should be emphasized.  First, the exact solution derived here is a result for the leading perturbative nilpotent branch of the reduced potential.  It determines the local boundary of the physical-volume trajectory, but by itself it does not establish a globally controlled compact solution.  Second, the closed formulae assume that the anti-D3 coefficient is effectively independent of \(\sigma\) along the local volume displacement.  When residual volume dependence in the Coulomb coefficient is important, the exact deformed equations of \Sec{sec:sigma_dependent} must be used.  Third, the Wilsonian analysis shows that the endpoint is sensitive to corrections because the curvature and the barrier height vanish.  Thus ordinary large-volume suppression must be supplemented by the local derivative inequalities controlling the displacement of the critical locus.  Fourth, the compact-flux filtration is a necessary arithmetic step, not a replacement for a full compactification analysis.

The results of this paper suggest several concrete directions for future work.

\begin{enumerate}[label=(\roman*)]

\item
\textit{Explicit compact period computations.}
The most direct continuation is to evaluate the map
\[
 N^I\longmapsto
 \big(\Wzero(N),g_s(N),M(N),K(N),\alpha(N),\beta(N),L(N)\big)
\]
in explicit orientifold or F-theory compactifications.  This requires the period vector, the flux superpotential, the tadpole pairing and the conifold expansion of the periods in the GKP/GVW framework and its F-theory lift \cite{Gukov:1999ya,Giddings:2001yu,Denef:2004dm,Denef:2005mm,Douglas:2006es,Grimm:2024exactFlux,McAllister:2024dS,Schachner:2025overview}.  The analytic formulas derived here then become a fast pre-filter for large flux scans.  A complete example would require solving \(D_AW=0\), computing the regular conifold periods, imposing the D3 tadpole, and deriving the perturbative K\"ahler coefficients from the compact geometry.

\item
\textit{Full K\"ahler-sector mass matrices.}
The present local analysis isolates the physical-volume direction.  In a genuine compact model the dangerous direction may be a mixture of the overall volume and orthogonal K\"ahler moduli.  The correct generalization is the physical mass matrix
\[
 {\cal M}^{I}{}_{J}=G^{IK}\nabla_K\nabla_JV
\]
and its nested Schur complement.  A systematic study of when the physical-volume fold is the first instability, rather than an orthogonal K\"ahler direction, is essential for compact applications.  This should be combined with the standard multifield effective-theory treatment of heavy-field integration and trajectory bending \cite{Gordon:2000hv,Langlois:2008qf,Tolley:2009fg,Achucarro:2010da,Achucarro:2012sm,Cespedes:2012hu,Burgess:2012dz,RenauxPetel:2015mga}.

\item
\textit{Numerical scans of the full necessary filter.}
The integer filtration shown in \Sec{sec:numerical_realization} can be extended to realistic ranges of \(M\), \(K\), \(L_{\eff}\), \(\Wzero\), and \(g_s\), in the spirit of flux-vacuum counting and explicit compact-throat searches \cite{Bousso:2000xa,Douglas:2003um,Ashok:2003gk,Denef:2004dm,Saltman:2004sn,Douglas:2006es,Blumenhagen:2006ci}.  Such scans should record not only whether a point survives, but also the distance to the critical locus, the remaining tadpole margin, the conifold-control margin, the Schur-control scale, and the scalar-amplitude normalization.  The natural output would be a ranked list of compact candidates for subsequent period and K\"ahler-sector analysis.

\item
\textit{Residual Coulomb-volume dependence.}
The plateau approximation is controlled by
\[
 |a_\sigma|\sigma\ll a,
 \qquad
 |a_{\sigma\sigma}|\sigma^2\ll a .
\]
It would be useful to compute \(a(\phi,\sigma)\) directly in warped compact geometries and then solve the deformed critical equations
\[
 F+\sigma^{7/2}a_\sigma=0,
 \qquad
 F_\sigma+\sigma^{7/2}a_{\sigma\sigma}
 +\frac32\sigma^{5/2}a_\sigma=0 .
\]
This would determine how far the closed branch remains quantitatively accurate in realistic throat embeddings of D3--anti-D3 dynamics \cite{Kachru:2003sx,McAllister:2008hb,Baumann:2014nda,Cicoli:2025origins}.

\item
\textit{Higher-order Wilsonian corrections.}
Near the critical locus, a correction small in the scalar potential can dominate the curvature or the barrier height.  Future work should evaluate the leading BBHL, loop, logarithmic, curvature, warping, and open-string threshold corrections in explicit compact models \cite{Becker:2002nn,Berg:2005ja,vonGersdorff:2005bf,Berg:2007wt,Cicoli:2007xp,Cicoli:2008gp,Conlon:2008cj,Grimm:2017pid,Grimm:2017okk,Antoniadis:2018hqy,Antoniadis:2019rkh} and test the local inequalities
\[
 |\partial_\chi^2\Delta V|\ll m_\chi^2,
 \qquad
 |\Delta V(\chi_+)-\Delta V(\chi_-)|\ll \Delta V_B .
\]
This would turn the formal Wilsonian response formulas into explicit control criteria.

\item
\textit{Multifield dynamics near the degeneration locus.}
The Schur-complement pole signals the breakdown of the static adiabatic description.  A dynamical multifield treatment should follow the coupled perturbations in the region where \(m_\chi^2\) becomes comparable to the Hubble scale, determine whether the trajectory bends strongly before the static endpoint, and compute the resulting corrections to the scalar power spectrum, non-Gaussianity, and isocurvature transfer \cite{Gordon:2000hv,Langlois:2008qf,Tolley:2009fg,Achucarro:2010da,Peterson:2010np,Gong:2011uw,Achucarro:2012sm,Cespedes:2012hu,Shiu:2011qw,RenauxPetel:2015mga}.

\item
\textit{Tunneling and finite-temperature stability.}
Because the barrier height scales as \(\Delta V_B\sim\mu^{3/2}\), the semiclassical decay rate of the K\"ahler vacuum should have a characteristic scaling near the endpoint.  Computing the Coleman--De Luccia or Hawking--Moss actions in the local normal form would give an analytic estimate of the lifetime of near-critical trajectories \cite{Coleman:1977py,Callan:1977pt,Coleman:1980aw,Hawking:1981fz,Parke:1982pm}.  This may be important for distinguishing trajectories that are locally metastable from those that are cosmologically long-lived.

\item
\textit{Endpoint statistics in the flux ensemble.}
The coordinate \(y\) converts the local branch into an elementary interval.  This makes it possible to study the distribution of distances to the critical endpoint in flux ensembles.  One can ask whether compact flux statistics preferentially populate small \(1-y\), where the admissible anti-D3 tension and the Wilsonian control margins are small, or whether typical controlled points lie well inside the branch.  This question naturally connects the present analytic branch coordinate to the statistical distribution of flux vacua and warped-throat data \cite{Bousso:2000xa,Douglas:2003um,Ashok:2003gk,Denef:2004dm,Denef:2005mm,Douglas:2006es}.

\item
\textit{Connection to CMB normalization.}
The inverse-quartic regime gives
\[
 n_s\simeq1-\frac{5}{3N},
 \qquad
 r\ll1,
\]
but the amplitude normalization must be compatible with the critical-tension bound, warped hierarchy, and Schur-control scale.  A full phenomenological scan should impose the scalar amplitude, the number of \(e\)-folds, the K\"ahler-valley existence condition, and the multifield control condition simultaneously.  This connects the local analytic bound to current CMB constraints and to the perturbative brane--antibrane construction \cite{Planck:2018jri,BICEP:2021xfz,Baumann:2014nda,Cicoli:2025origins}.

\item
\textit{General weighted positive-energy terms.}
The appendices show how the algebra changes for more general volume weights.  This can be used to compare different anti-D3 descriptions, alternative nilpotent branches, D-term-like positive-energy contributions, or other perturbative string-inflation potentials \cite{Volkov:1973ix,Rocek:1978nb,Ferrara:2013rsa,Bergshoeff:2015jxa,Kallosh:2015nia,Vercnocke:2016fbt,Binetruy:1996xj,Halyo:1996pp,Burgess:2003ic}.  The central question is whether equally simple uniformizing coordinates exist beyond the \(\sigma^{-2}\) term considered in the main text.

\end{enumerate}

The central lesson is that perturbative brane--antibrane inflation contains an analytic, model-independent local boundary for the physical-volume trajectory once the reduced potential takes the form \Eq{eq:U_def_intro}.  The critical anti-D3 tension, the K\"ahler mass collapse, the barrier-height scaling, and the Schur-complement singularity are different manifestations of the same saddle-node degeneration.  Compact data must therefore satisfy not only the usual flux, tadpole, hierarchy and CMB constraints, but also the exact K\"ahler-valley existence condition derived here.  This condition is simple enough to be imposed analytically and restrictive enough to provide a useful guide for future compact searches.

\appendix

% ============================================================
\section{Derivative identities at the degeneration locus}\label{app:derivatives}
% ============================================================

For
\begin{equation}\label{app:U_def}
 U=a\sigma^{-2}+\alpha\sigma^{-4}-\beta\sigma^{-9/2}(\ln\sigma-L),
\end{equation}
one has
\begin{align}\label{app:derivs_U}
 U_\sigma&=\sigma^{-11/2}F,\nn\\
 U_{\sigma\sigma}&=-\frac{11}{2}\sigma^{-13/2}F+\sigma^{-11/2}F_\sigma,\nn\\
 U_{\sigma\sigma\sigma}&=\frac{143}{4}\sigma^{-15/2}F
 -11\sigma^{-13/2}F_\sigma+\sigma^{-11/2}F_{\sigma\sigma}.
\end{align}
At the degeneration point \(F=F_\sigma=0\),
\begin{equation}\label{app:Usss}
 U_{\sigma\sigma\sigma,c}=\sigma_c^{-11/2}F_{\sigma\sigma,c}.
\end{equation}
The canonical cubic coefficient follows from \(\dd\sigma/\dd\chi=\sqrt{2/3}\sigma\):
\begin{equation}\label{app:canonical_cubic}
 V_{\chi\chi\chi,c}=3|\Wzero|^2
 \left(\frac{\dd\sigma}{\dd\chi}\right)^3U_{\sigma\sigma\sigma,c}
 =2\sqrt{\frac23}|\Wzero|^2\sigma_c^{-5/2}F_{\sigma\sigma,c}.
\end{equation}
On the elementary branch,
\begin{align}\label{app:F_derivs_y}
 F_{\sigma\sigma,c}&=-\frac{9\beta}{4}\sigma_c^{-2}\Xi,\nn\\
 F_{\sigma\sigma\sigma,c}&=\frac{45\beta}{8}\sigma_c^{-3},\nn\\
 \Xi&=5-4y.
\end{align}

% ============================================================
\section{General weighted anti-D3 terms}\label{app:general_weight}
% ============================================================

For
\begin{equation}\label{app:weighted_U}
 U=A w(\sigma)+P(\sigma),
\end{equation}
the degeneration equation is
\begin{equation}\label{app:weighted_eq}
 P''w'-P'w''=0.
\end{equation}
If \(w=\sigma^{-n}\), this becomes
\begin{equation}\label{app:weighted_power}
 \sigma P''+(n+1)P'=0,
\end{equation}
and
\begin{equation}\label{app:weighted_Ac}
 A_c=\frac{\sigma^{n+1}}{n}P'.
\end{equation}
The leading anti-D3 term corresponds to \(n=2\), giving \(\sigma P''+3P'=0\) and \(a_c=\sigma^3P'/2\).

% ============================================================
\section{Endpoint algebra}\label{app:endpoint}
% ============================================================

With \(s=1-y\),
\begin{equation}\label{app:theta_s}
 \theta=(1-s)\ee^{4s/5}.
\end{equation}
The logarithm is
\begin{equation}\label{app:logtheta_s}
 \ln\theta=\ln(1-s)+\frac45s
 =-\frac15s-\frac12s^2-\frac13s^3-\frac14s^4-\cdots .
\end{equation}
Exponentiating gives
\begin{equation}\label{app:theta_series}
 \theta=1-\frac15s-\frac{12}{25}s^2-\frac{88}{375}s^3+\order(s^4).
\end{equation}
Since \(a_c=(9\beta/10)\ee^{-5L/2-14/9-4(1-s)}s\),
\begin{equation}\label{app:ac_s}
 a_c=\frac{9\beta}{10}\ee^{-5L/2-50/9}s\left(1+4s+8s^2+\frac{32}{3}s^3+\cdots\right).
\end{equation}
Combining this with the inverse series for \(s(\delta)\), \(\delta=1-\theta\), yields \Eq{eq:ac_delta}.

% ============================================================
\section{Flux-interval sensitivities}\label{app:flux_sensitivity}
% ============================================================

The boundary equation at fixed \((\rho,L,\beta,|W_0|)\) is
\begin{equation}\label{app:boundary_eq}
 a_0=C_a\ee^{-4y}(1-y),
 \qquad
 C_a=\frac{9\beta}{10}\ee^{-5L/2-14/9}.
\end{equation}
Defining \(Z=4\ee^4a_0/C_a\), the solution is
\begin{equation}\label{app:ymax}
 y_{\rm max}=1-\frac14\LambertW_0(Z).
\end{equation}
For any parameter \(p\),
\begin{equation}\label{app:sensitivity_general}
 \frac{\partial y_{\rm max}}{\partial\ln p}
 =-\frac14\frac{\LambertW_0(Z)}{1+\LambertW_0(Z)}
 \frac{\partial\ln Z}{\partial\ln p}.
\end{equation}
Since \(Z\propto\ee^{-4\rho}|W_0|^{-2}\beta^{-1}\ee^{5L/2}\),
\begin{subequations}\label{app:sensitivities}
\begin{align}
 \frac{\partial y_{\rm max}}{\partial\rho}
 &=\frac{\LambertW_0(Z)}{1+\LambertW_0(Z)},\label{app:dy_drho}\\
 \frac{\partial y_{\rm max}}{\partial\ln|W_0|}
 &=\frac12\frac{\LambertW_0(Z)}{1+\LambertW_0(Z)},\label{app:dy_dW0}\\
 \frac{\partial y_{\rm max}}{\partial\ln\beta}
 &=\frac14\frac{\LambertW_0(Z)}{1+\LambertW_0(Z)},\label{app:dy_dbeta}\\
 \frac{\partial y_{\rm max}}{\partial L}
 &=-\frac58\frac{\LambertW_0(Z)}{1+\LambertW_0(Z)}.
 \label{app:dy_dL}
\end{align}
\end{subequations}
The sign of \Eq{app:dy_dW0} is positive: increasing \(|W_0|\) reduces the plateau tension parameter \(a_0\) and enlarges the interval.

% ============================================================
\section{Nested Schur complements}\label{app:nested_schur}
% ============================================================

Let \(\Phi^I=(\phi,\sigma,\psi^a)\) be the scalar fields, with field-space metric \(G_{IJ}\), and let \(E_A{}^I\) be a local orthonormal vielbein.  The physical Hessian is
\begin{equation}\label{app:physical_hessian}
 {\cal H}_{AB}=E_A{}^IE_B{}^J\nabla_I\nabla_JV.
\end{equation}
In the orthonormal basis \((\phi,\chi,a)\), where \(a\) labels the orthogonal K\"ahler directions, write
\begin{equation}\label{app:full_H}
 {\cal H}=\begin{pmatrix}
 {\cal H}_{\phi\phi} & {\cal H}_{\phi\chi} & {\cal H}_{\phi b}\\
 {\cal H}_{\chi\phi} & {\cal H}_{\chi\chi} & {\cal H}_{\chi b}\\
 {\cal H}_{a\phi} & {\cal H}_{a\chi} & {\cal H}_{ab}
 \end{pmatrix}.
\end{equation}
Integrating out the orthogonal K\"ahler fields first gives
\begin{align}\label{app:reduced_blocks}
 \widehat{\cal H}_{\phi\phi}
 &={\cal H}_{\phi\phi}-{\cal H}_{\phi a}({\cal H}^{-1})^{ab}{\cal H}_{b\phi},\nn\\
 \widehat{\cal H}_{\phi\chi}
 &={\cal H}_{\phi\chi}-{\cal H}_{\phi a}({\cal H}^{-1})^{ab}{\cal H}_{b\chi},\nn\\
 \widehat{\cal H}_{\chi\chi}
 &={\cal H}_{\chi\chi}-{\cal H}_{\chi a}({\cal H}^{-1})^{ab}{\cal H}_{b\chi}.
\end{align}
The final adiabatic Hessian is
\begin{equation}\label{app:nested_final}
 {\cal H}_{\phi\phi}^{\rm ad}
 =\widehat{\cal H}_{\phi\phi}
 -\frac{\widehat{\cal H}_{\phi\chi}^2}{\widehat{\cal H}_{\chi\chi}}.
\end{equation}
The one-volume expression in \Sec{sec:schur} is recovered when the orthogonal corrections in \Eq{app:reduced_blocks} are negligible.  If \(\widehat{\cal H}_{\chi\chi}\) vanishes before \({\cal H}_{\chi\chi}\), the compact theory degenerates earlier than predicted by the one-volume potential.

% ============================================================
\section{CMB relations in the inverse-quartic regime}\label{app:cmb}
% ============================================================

For
\begin{equation}\label{app:V_inv}
 V(\phi)=V_0\left(1-\frac{C}{\phi^4}\right),
\end{equation}
with \(C/\phi^4\ll1\),
\begin{equation}\label{app:Vprime}
 \frac{V'}{V}\simeq\frac{4C}{\phi^5},
 \qquad
 \frac{V''}{V}\simeq-\frac{20C}{\phi^6}.
\end{equation}
Then
\begin{equation}\label{app:N}
 N=\int_{\phi_{\rm end}}^{\phi_*}\frac{V}{V'}\dd\phi
 \simeq\frac{\phi_*^6}{24C},
\end{equation}
when \(\phi_*\gg\phi_{\rm end}\).  The scalar amplitude is
\begin{equation}\label{app:As}
 \Aobs=\frac{V_*}{24\pi^2\epsilon_*}
 \simeq\frac{V_0}{12\pi^2}\frac{\phi_*^{10}}{16C^2}
 =\frac{V_0(24N_*)^{5/3}}{192\pi^2C^{1/3}}.
\end{equation}
The spectral index is
\begin{equation}\label{app:ns}
 n_s=1-6\epsilon_*+2\eta_*
 \simeq1-\frac{5}{3N_*},
\end{equation}
and \(r=16\epsilon_*\) is parametrically small for the same hierarchy.

\bibliography{StringIRefs}

@article{Giddings:2001yu,
  author        = {Giddings, Steven B. and Kachru, Shamit and Polchinski, Joseph},
  title         = {Hierarchies from fluxes in string compactifications},
  journal       = {Phys. Rev. D},
  volume        = {66},
  pages         = {106006},
  year          = {2002},
  doi           = {10.1103/PhysRevD.66.106006},
  eprint        = {hep-th/0105097},
  archivePrefix = {arXiv}
}

@article{Coleman:1977py,
  author        = {Coleman, Sidney R.},
  title         = {{The Fate of the False Vacuum. 1. Semiclassical Theory}},
  reportNumber  = {HUTP-77/A004},
  doi           = {10.1103/PhysRevD.15.2929},
  journal       = {Phys. Rev. D},
  volume        = {15},
  pages         = {2929--2936},
  year          = {1977},
  note          = {[Erratum: Phys. Rev. D 16, 1248 (1977)]}
}

@article{Callan:1977pt,
  author        = {Callan, Curtis G. and Coleman, Sidney R.},
  title         = {{The Fate of the False Vacuum. 2. First Quantum Corrections}},
  reportNumber  = {HUTP-77/A032},
  doi           = {10.1103/PhysRevD.16.1762},
  journal       = {Phys. Rev. D},
  volume        = {16},
  pages         = {1762--1768},
  year          = {1977}
}

@article{Coleman:1980aw,
  author        = {Coleman, Sidney R. and De Luccia, Frank},
  title         = {{Gravitational Effects on and of Vacuum Decay}},
  reportNumber  = {HUTP-80/A001},
  doi           = {10.1103/PhysRevD.21.3305},
  journal       = {Phys. Rev. D},
  volume        = {21},
  pages         = {3305},
  year          = {1980}
}

@article{Hawking:1981fz,
  author        = {Hawking, S. W. and Moss, I. G.},
  title         = {{Supercooled Phase Transitions in the Very Early Universe}},
  reportNumber  = {PRINT-82-0022 (CAMBRIDGE)},
  doi           = {10.1016/0370-2693(82)90946-7},
  journal       = {Phys. Lett. B},
  volume        = {110},
  pages         = {35--38},
  year          = {1982}
}

@article{Parke:1982pm,
  author        = {Parke, Stephen J.},
  title         = {{Gravity, the Decay of the False Vacuum and the New Inflationary Universe Scenario}},
  reportNumber  = {Print-82-0695 (FERMILAB)},
  doi           = {10.1016/0370-2693(83)90648-X},
  journal       = {Phys. Lett. B},
  volume        = {121},
  pages         = {313--315},
  year          = {1983}
}

@article{Binetruy:1996xj,
  author        = {Binetruy, Pierre and Dvali, Gia R.},
  title         = {{D-term inflation}},
  eprint        = {hep-ph/9606342},
  archivePrefix = {arXiv},
  reportNumber  = {CERN-TH-96-149, LPTHE-ORSAY-96-60, IFUP-TH-32-96},
  doi           = {10.1016/0370-2693(96)01083-0},
  journal       = {Phys. Lett. B},
  volume        = {388},
  pages         = {241--246},
  year          = {1996}
}

@article{Halyo:1996pp,
  author        = {Halyo, Edi},
  title         = {{Hybrid inflation from supergravity D-terms}},
  eprint        = {hep-ph/9606423},
  archivePrefix = {arXiv},
  reportNumber  = {SU-ITP-96-25},
  doi           = {10.1016/0370-2693(96)01086-6},
  journal       = {Phys. Lett. B},
  volume        = {387},
  pages         = {43--47},
  year          = {1996}
}

@article{Burgess:2003ic,
  author        = {Burgess, C. P. and Kallosh, Renata and Quevedo, Fernando},
  title         = {{de Sitter string vacua from supersymmetric D-terms}},
  eprint        = {hep-th/0309187},
  archivePrefix = {arXiv},
  reportNumber  = {OUTP-03-32P, SU-ITP-03-26},
  doi           = {10.1088/1126-6708/2003/10/056},
  journal       = {JHEP},
  volume        = {10},
  pages         = {056},
  year          = {2003}
}

@article{Gukov:1999ya,
  author        = {Gukov, Sergei and Vafa, Cumrun and Witten, Edward},
  title         = {CFT's from Calabi--Yau four-folds},
  journal       = {Nucl. Phys. B},
  volume        = {584},
  pages         = {69--108},
  year          = {2000},
  doi           = {10.1016/S0550-3213(00)00373-4},
  eprint        = {hep-th/9906070},
  archivePrefix = {arXiv}
}

@article{Klebanov:1998hh,
  author={Klebanov, Igor R. and Witten, Edward},
  title={Superconformal field theory on three-branes at a Calabi-Yau singularity},
  journal={Nucl. Phys. B},
  volume={536},
  pages={199--218},
  year={1998},
  doi={10.1016/S0550-3213(98)00654-3},
  eprint={hep-th/9807080},
  archivePrefix={arXiv}
}

@article{Klebanov:2000nc,
  author={Klebanov, Igor R. and Tseytlin, Arkady A.},
  title={Gravity duals of supersymmetric SU(N) x SU(N+M) gauge theories},
  journal={Nucl. Phys. B},
  volume={578},
  pages={123--138},
  year={2000},
  doi={10.1016/S0550-3213(00)00206-6},
  eprint={hep-th/0002159},
  archivePrefix={arXiv}
}

@article{Klebanov:2000hb,
  author        = {Klebanov, Igor R. and Strassler, Matthew J.},
  title         = {Supergravity and a confining gauge theory: Duality cascades and chiSB-resolution of naked singularities},
  journal       = {JHEP},
  volume        = {08},
  pages         = {052},
  year          = {2000},
  doi           = {10.1088/1126-6708/2000/08/052},
  eprint        = {hep-th/0007191},
  archivePrefix = {arXiv}
}

@article{Grana:2005jc,
  author={Grana, Mariana},
  title={Flux compactifications in string theory: A comprehensive review},
  journal={Phys. Rept.},
  volume={423},
  pages={91--158},
  year={2006},
  doi={10.1016/j.physrep.2005.10.008},
  eprint={hep-th/0509003},
  archivePrefix={arXiv}
}

@article{DeWolfe:2002nn,
  author={DeWolfe, Oliver and Giddings, Steven B.},
  title={Scales and hierarchies in warped compactifications and brane worlds},
  journal={Phys. Rev. D},
  volume={67},
  pages={066008},
  year={2003},
  doi={10.1103/PhysRevD.67.066008},
  eprint={hep-th/0208123},
  archivePrefix={arXiv}
}

@article{Shiu:2008ry,
  author={Shiu, Gary and Underwood, Bret},
  title={Observing the Geometry of Warped Compactification via Cosmic Inflation},
  journal={Phys. Rev. Lett.},
  volume={98},
  pages={051301},
  year={2007},
  doi={10.1103/PhysRevLett.98.051301},
  eprint={hep-th/0610151},
  archivePrefix={arXiv}
}

@article{Denef:2008wq,
  author={Denef, Frederik},
  title={Les Houches Lectures on Constructing String Vacua},
  journal={Les Houches},
  volume={87},
  pages={483--610},
  year={2008},
  eprint={0803.1194},
  archivePrefix={arXiv},
  primaryClass={hep-th}
}

@article{Dvali:1998pa,
  author        = {Dvali, G. R. and Tye, S. H. Henry},
  title         = {Brane inflation},
  journal       = {Phys. Lett. B},
  volume        = {450},
  pages         = {72--82},
  year          = {1999},
  doi           = {10.1016/S0370-2693(99)00132-X},
  eprint        = {hep-ph/9812483},
  archivePrefix = {arXiv}
}

@article{Burgess:2001fx,
  author        = {Burgess, C. P. and Majumdar, M. and Nolte, D. and Quevedo, F. and Rajesh, G. and Zhang, R. J.},
  title         = {The inflationary brane-antibrane universe},
  journal       = {JHEP},
  volume        = {07},
  pages         = {047},
  year          = {2001},
  doi           = {10.1088/1126-6708/2001/07/047},
  eprint        = {hep-th/0105204},
  archivePrefix = {arXiv}
}

@article{Alexander:2001ks,
  author        = {Alexander, Stephon H. S.},
  title         = {Inflation from D--anti-D brane annihilation},
  journal       = {Phys. Rev. D},
  volume        = {65},
  pages         = {023507},
  year          = {2002},
  doi           = {10.1103/PhysRevD.65.023507},
  eprint        = {hep-th/0105032},
  archivePrefix = {arXiv}
}

@article{GarciaBellido:2001ky,
  author        = {Garcia-Bellido, Juan and Rabadan, Raul and Zamora, Fernando},
  title         = {Inflationary scenarios from branes at angles},
  journal       = {JHEP},
  volume        = {01},
  pages         = {036},
  year          = {2002},
  doi           = {10.1088/1126-6708/2002/01/036},
  eprint        = {hep-th/0112147},
  archivePrefix = {arXiv}
}

@article{Kachru:2003sx,
  author        = {Kachru, Shamit and Kallosh, Renata and Linde, Andrei D. and Maldacena, Juan Martin and McAllister, Liam and Trivedi, Sandip P.},
  title         = {Towards inflation in string theory},
  journal       = {JCAP},
  volume        = {10},
  pages         = {013},
  year          = {2003},
  doi           = {10.1088/1475-7516/2003/10/013},
  eprint        = {hep-th/0308055},
  archivePrefix = {arXiv}
}

@article{Silverstein:2003hf,
  author={Silverstein, Eva and Tong, David},
  title={Scalar speed limits and cosmology: Acceleration from D-cceleration},
  journal={Phys. Rev. D},
  volume={70},
  pages={103505},
  year={2004},
  doi={10.1103/PhysRevD.70.103505},
  eprint={hep-th/0310221},
  archivePrefix={arXiv}
}

@article{Alishahiha:2004eh,
  author={Alishahiha, Mohsen and Silverstein, Eva and Tong, David},
  title={DBI in the sky},
  journal={Phys. Rev. D},
  volume={70},
  pages={123505},
  year={2004},
  doi={10.1103/PhysRevD.70.123505},
  eprint={hep-th/0404084},
  archivePrefix={arXiv}
}

@article{Firouzjahi:2005dh,
  author={Firouzjahi, Hassan and Tye, S. H. Henry},
  title={Closer towards inflation in string theory},
  journal={Phys. Lett. B},
  volume={584},
  pages={147--154},
  year={2004},
  doi={10.1016/j.physletb.2004.01.062},
  eprint={hep-th/0312020},
  archivePrefix={arXiv}
}

@article{Chen:2005ad,
  author={Chen, Xingang},
  title={Inflation from warped space},
  journal={JHEP},
  volume={08},
  pages={045},
  year={2005},
  doi={10.1088/1126-6708/2005/08/045},
  eprint={hep-th/0501184},
  archivePrefix={arXiv}
}

@article{Chen:2006nt,
  author={Chen, Xingang},
  title={Running non-Gaussianities in DBI inflation},
  journal={Phys. Rev. D},
  volume={72},
  pages={123518},
  year={2005},
  doi={10.1103/PhysRevD.72.123518},
  eprint={astro-ph/0507053},
  archivePrefix={arXiv}
}

@article{Lidsey:2007gq,
  author={Lidsey, James E. and Huston, Ian and Thomas, Steven and Ward, John},
  title={Towards observational constraints on DBI inflation},
  journal={JCAP},
  volume={02},
  pages={002},
  year={2007},
  doi={10.1088/1475-7516/2007/02/002},
  eprint={hep-th/0604141},
  archivePrefix={arXiv}
}

@article{Krause:2007jr,
  author={Krause, Axel and Pajer, Enrico},
  title={Chasing Brane Inflation in String-Theory},
  journal={JCAP},
  volume={07},
  pages={023},
  year={2008},
  doi={10.1088/1475-7516/2008/07/023},
  eprint={0705.4682},
  archivePrefix={arXiv},
  primaryClass={hep-th}
}

@article{Pajer:2008uy,
  author={Pajer, Enrico},
  title={Inflation at the Tip},
  journal={JCAP},
  volume={04},
  pages={031},
  year={2008},
  doi={10.1088/1475-7516/2008/04/031},
  eprint={0802.2916},
  archivePrefix={arXiv},
  primaryClass={hep-th}
}

@article{Chen:2008hz,
  author={Chen, Xingang},
  title={Fine-Tuning in DBI Inflationary Mechanism},
  journal={JCAP},
  volume={12},
  pages={009},
  year={2008},
  doi={10.1088/1475-7516/2008/12/009},
  eprint={0807.3191},
  archivePrefix={arXiv},
  primaryClass={hep-th}
}

@article{McAllister:2008hb,
  author={McAllister, Liam and Silverstein, Eva and Westphal, Alexander},
  title={Gravity Waves and Linear Inflation from Axion Monodromy},
  journal={Phys. Rev. D},
  volume={82},
  pages={046003},
  year={2010},
  doi={10.1103/PhysRevD.82.046003},
  eprint={0808.0706},
  archivePrefix={arXiv},
  primaryClass={hep-th}
}

@book{Baumann:2014nda,
  author = {Baumann, Daniel and McAllister, Liam},
  title = {Inflation and String Theory},
  publisher = {Cambridge University Press},
  year = {2015},
  eprint = {1404.2601},
  archivePrefix = {arXiv},
  primaryClass = {hep-th}
}

@article{Kachru:2003aw,
  author        = {Kachru, Shamit and Kallosh, Renata and Linde, Andrei D. and Trivedi, Sandip P.},
  title         = {De Sitter vacua in string theory},
  journal       = {Phys. Rev. D},
  volume        = {68},
  pages         = {046005},
  year          = {2003},
  doi           = {10.1103/PhysRevD.68.046005},
  eprint        = {hep-th/0301240},
  archivePrefix = {arXiv}
}

@article{Baumann:2006th,
  author        = {Baumann, Daniel and Dymarsky, Anatoly and Klebanov, Igor R. and Maldacena, Juan Martin and McAllister, Liam and Murugan, Arvind},
  title         = {On D3-brane potentials in compactifications with fluxes and wrapped D-branes},
  journal       = {JHEP},
  volume        = {11},
  pages         = {031},
  year          = {2006},
  doi           = {10.1088/1126-6708/2006/11/031},
  eprint        = {hep-th/0607050},
  archivePrefix = {arXiv}
}

@article{Baumann:2007ah,
  author        = {Baumann, Daniel and Dymarsky, Anatoly and Klebanov, Igor R. and McAllister, Liam},
  title         = {Towards an explicit model of D-brane inflation},
  journal       = {JCAP},
  volume        = {01},
  pages         = {024},
  year          = {2008},
  doi           = {10.1088/1475-7516/2008/01/024},
  eprint        = {0706.0360},
  archivePrefix = {arXiv},
  primaryClass  = {hep-th}
}

@article{Baumann:2009ni,
  author        = {Baumann, Daniel and McAllister, Liam},
  title         = {Advances in inflation in string theory},
  journal       = {Ann. Rev. Nucl. Part. Sci.},
  volume        = {59},
  pages         = {67--94},
  year          = {2009},
  doi           = {10.1146/annurev.nucl.010909.083524},
  eprint        = {0901.0265},
  archivePrefix = {arXiv},
  primaryClass  = {hep-th}
}

@article{McAllister:2007bg,
  author        = {McAllister, Liam and Silverstein, Eva},
  title         = {String cosmology: A review},
  journal       = {Gen. Rel. Grav.},
  volume        = {40},
  pages         = {565--605},
  year          = {2008},
  doi           = {10.1007/s10714-007-0556-6},
  eprint        = {0710.2951},
  archivePrefix = {arXiv},
  primaryClass  = {hep-th}
}

@article{Cicoli:2025origins,
  author        = {Cicoli, Michele and Hughes, Christopher and Kamal, Ahmed Rakin and Marino, Francesco and Quevedo, Fernando and Ramos-Hamud, Mario and Villa, Gonzalo},
  title         = {Back to the origins of brane--antibrane inflation},
  journal       = {Eur. Phys. J. C},
  volume        = {85},
  number        = {3},
  pages         = {315},
  year          = {2025},
  doi           = {10.1140/epjc/s10052-025-13982-9},
  eprint        = {2410.00097},
  archivePrefix = {arXiv},
  primaryClass  = {hep-th}
}

@article{Planck:2018jri,
  author        = {Akrami, Y. and others},
  collaboration = {Planck},
  title         = {Planck 2018 results. X. Constraints on inflation},
  journal       = {Astron. Astrophys.},
  volume        = {641},
  pages         = {A10},
  year          = {2020},
  doi           = {10.1051/0004-6361/201833887},
  eprint        = {1807.06211},
  archivePrefix = {arXiv},
  primaryClass  = {astro-ph.CO}
}

@article{BICEP:2021xfz,
  author        = {Ade, P. A. R. and others},
  collaboration = {BICEP/Keck},
  title         = {Improved constraints on primordial gravitational waves using Planck, WMAP, and BICEP/Keck observations through the 2018 observing season},
  journal       = {Phys. Rev. Lett.},
  volume        = {127},
  number        = {15},
  pages         = {151301},
  year          = {2021},
  doi           = {10.1103/PhysRevLett.127.151301},
  eprint        = {2110.00483},
  archivePrefix = {arXiv},
  primaryClass  = {astro-ph.CO}
}

@article{Becker:2002nn,
  author        = {Becker, Katrin and Becker, Melanie and Haack, Michael and Louis, Jan},
  title         = {Supersymmetry breaking and alpha-prime corrections to flux induced potentials},
  journal       = {JHEP},
  volume        = {06},
  pages         = {060},
  year          = {2002},
  doi           = {10.1088/1126-6708/2002/06/060},
  eprint        = {hep-th/0204254},
  archivePrefix = {arXiv}
}

@article{Volkov:1973ix,
  author        = {Volkov, D. V. and Akulov, V. P.},
  title         = {Is the neutrino a Goldstone particle?},
  journal       = {Phys. Lett. B},
  volume        = {46},
  pages         = {109--110},
  year          = {1973},
  doi           = {10.1016/0370-2693(73)90490-5}
}

@article{Rocek:1978nb,
  author        = {Rocek, M.},
  title         = {Linearizing the Volkov--Akulov model},
  journal       = {Phys. Rev. Lett.},
  volume        = {41},
  pages         = {451--453},
  year          = {1978},
  doi           = {10.1103/PhysRevLett.41.451}
}

@article{Lindstrom:1979kq,
  author        = {Lindstrom, Ulf and Rocek, Martin},
  title         = {Constrained local superfields},
  journal       = {Phys. Rev. D},
  volume        = {19},
  pages         = {2300--2303},
  year          = {1979},
  doi           = {10.1103/PhysRevD.19.2300}
}

@article{Ferrara:2013rsa,
  author={Ferrara, Sergio and Kallosh, Renata and Linde, Andrei and Porrati, Massimo},
  title={Minimal Supergravity Models of Inflation},
  journal={Phys. Rev. D},
  volume={88},
  pages={085038},
  year={2013},
  doi={10.1103/PhysRevD.88.085038},
  eprint={1307.7696},
  archivePrefix={arXiv},
  primaryClass={hep-th}
}

@article{Bergshoeff:2015jxa,
  author        = {Bergshoeff, Eric A. and Dasgupta, Keshav and Kallosh, Renata and Van Proeyen, Antoine and Wrase, Timm},
  title         = {D3 and dS},
  journal       = {JHEP},
  volume        = {05},
  pages         = {058},
  year          = {2015},
  doi           = {10.1007/JHEP05(2015)058},
  eprint        = {1502.07627},
  archivePrefix = {arXiv},
  primaryClass  = {hep-th}
}

@article{Garcia-Etxebarria:2015lif,
  author        = {Garcia-Etxebarria, Inaki and Kallosh, Renata and Quevedo, Fernando and Uranga, Angel M. and Valenzuela, Irene},
  title         = {The nilpotent goldstino and anti-D3 branes in string theory},
  journal       = {JHEP},
  volume        = {12},
  pages         = {178},
  year          = {2015},
  doi           = {10.1007/JHEP12(2015)178},
  eprint        = {1510.04227},
  archivePrefix = {arXiv},
  primaryClass  = {hep-th}
}

@article{DallAgata:2015zxp,
  author={Dall'Agata, Gianguido and Farakos, Fotis},
  title={Constrained superfields in Supergravity},
  journal={JHEP},
  volume={02},
  pages={101},
  year={2016},
  doi={10.1007/JHEP02(2016)101},
  eprint={1512.02158},
  archivePrefix={arXiv},
  primaryClass={hep-th}
}

@article{Kallosh:2015nia,
  author={Kallosh, Renata and Linde, Andrei and Wrase, Timm},
  title={Coupling the inflationary sector to matter},
  journal={JHEP},
  volume={04},
  pages={027},
  year={2016},
  doi={10.1007/JHEP04(2016)027},
  eprint={1601.00005},
  archivePrefix={arXiv},
  primaryClass={hep-th}
}

@article{Kallosh:2015tea,
  author={Kallosh, Renata and Quevedo, Fernando and Uranga, Angel M.},
  title={String Theory Realizations of the Nilpotent Goldstino},
  journal={JHEP},
  volume={12},
  pages={039},
  year={2015},
  doi={10.1007/JHEP12(2015)039},
  eprint={1507.07556},
  archivePrefix={arXiv},
  primaryClass={hep-th}
}

@article{Vercnocke:2016fbt,
  author        = {Vercnocke, Bert and Wrase, Timm},
  title         = {Constrained superfields from an anti-D3-brane in KKLT},
  journal       = {JHEP},
  volume        = {08},
  pages         = {132},
  year          = {2016},
  doi           = {10.1007/JHEP08(2016)132},
  eprint        = {1605.03961},
  archivePrefix = {arXiv},
  primaryClass  = {hep-th}
}

@article{McDonough:2016der,
  author={McDonough, Evan and Scalisi, Marco},
  title={Inflation from Nilpotent Kahler Corrections},
  journal={JCAP},
  volume={11},
  pages={028},
  year={2016},
  doi={10.1088/1475-7516/2016/11/028},
  eprint={1609.00364},
  archivePrefix={arXiv},
  primaryClass={hep-th}
}

@article{Berg:2005ja,
  author        = {Berg, Marcus and Haack, Michael and Kors, Boris},
  title         = {String loop corrections to Kähler potentials in orientifolds},
  journal       = {JHEP},
  volume        = {11},
  pages         = {030},
  year          = {2005},
  doi           = {10.1088/1126-6708/2005/11/030},
  eprint        = {hep-th/0508043},
  archivePrefix = {arXiv}
}

@article{vonGersdorff:2005bf,
  author = {von Gersdorff, Gero and Hebecker, Arthur},
  title = {Kahler corrections for the volume modulus of flux compactifications},
  journal = {Phys. Lett. B},
  volume = {624},
  pages = {270--274},
  year = {2005},
  eprint = {hep-th/0507131},
  archivePrefix = {arXiv}
}

@article{Berg:2007wt,
  author        = {Berg, Marcus and Haack, Michael and Pajer, Enrico},
  title         = {Jumping through loops: On soft terms from large volume compactifications},
  journal       = {JHEP},
  volume        = {09},
  pages         = {031},
  year          = {2007},
  doi           = {10.1088/1126-6708/2007/09/031},
  eprint        = {0704.0737},
  archivePrefix = {arXiv},
  primaryClass  = {hep-th}
}

@article{Cicoli:2007xp,
  author = {Cicoli, Michele and Conlon, Joseph P. and Quevedo, Fernando},
  title = {Systematics of string loop corrections in type IIB Calabi-Yau flux compactifications},
  journal = {JHEP},
  volume = {01},
  pages = {052},
  year = {2008},
  eprint = {0708.1873},
  archivePrefix = {arXiv},
  primaryClass = {hep-th}
}

@article{Cicoli:2008gp,
  author        = {Cicoli, Michele and Conlon, Joseph P. and Quevedo, Fernando},
  title         = {General analysis of LARGE volume scenarios with string loop moduli stabilisation},
  journal       = {JHEP},
  volume        = {10},
  pages         = {105},
  year          = {2008},
  doi           = {10.1088/1126-6708/2008/10/105},
  eprint        = {0805.1029},
  archivePrefix = {arXiv},
  primaryClass  = {hep-th}
}

@article{Conlon:2008cj,
  author={Conlon, Joseph P. and Quevedo, Fernando},
  title={Kahler moduli inflation},
  journal={JHEP},
  volume={01},
  pages={146},
  year={2006},
  doi={10.1088/1126-6708/2006/01/146},
  eprint={hep-th/0509012},
  archivePrefix={arXiv}
}

@article{Antoniadis:2018hqy,
  author        = {Antoniadis, Ignatios and Chen, Yifan and Leontaris, George K.},
  title         = {Logarithmic loop corrections, moduli stabilisation and de Sitter vacua in string theory},
  journal       = {JHEP},
  volume        = {01},
  pages         = {149},
  year          = {2020},
  doi           = {10.1007/JHEP01(2020)149},
  eprint        = {1909.10525},
  archivePrefix = {arXiv},
  primaryClass  = {hep-th}
}

@article{Antoniadis:2019rkh,
  author        = {Antoniadis, Ignatios and Chen, Yifan and Leontaris, George K.},
  title         = {Perturbative moduli stabilisation in type IIB/F-theory framework},
  journal       = {Eur. Phys. J. C},
  volume        = {78},
  number        = {9},
  pages         = {766},
  year          = {2018},
  doi           = {10.1140/epjc/s10052-018-6240-1},
  eprint        = {1803.08941},
  archivePrefix = {arXiv},
  primaryClass  = {hep-th}
}

@article{Grimm:2017pid,
  author        = {Grimm, Thomas W. and Savelli, Raffaele and Weissenbacher, Matthias},
  title         = {On alpha-prime corrections in N=1 F-theory compactifications},
  journal       = {Phys. Lett. B},
  volume        = {725},
  pages         = {431--436},
  year          = {2013},
  doi           = {10.1016/j.physletb.2013.07.050},
  eprint        = {1303.3317},
  archivePrefix = {arXiv},
  primaryClass  = {hep-th}
}

@article{Grimm:2017okk,
  author        = {Grimm, Thomas W. and Weissenbacher, Matthias},
  title         = {On Kähler coordinates and corrections in N=1 compactifications},
  journal       = {JHEP},
  volume        = {10},
  pages         = {039},
  year          = {2017},
  doi           = {10.1007/JHEP10(2017)039},
  eprint        = {1702.08404},
  archivePrefix = {arXiv},
  primaryClass  = {hep-th}
}

@article{Cicoli:2023opf,
  author        = {Cicoli, Michele and Licheri, Marco and Maharana, Anshuman and Quevedo, Fernando},
  title         = {A note on logarithmic corrections to the Kähler potential},
  journal       = {JHEP},
  volume        = {01},
  pages         = {032},
  year          = {2024},
  eprint        = {2309.11697},
  archivePrefix = {arXiv},
  primaryClass  = {hep-th}
}

@article{Cicoli:2024fva,
  author        = {Cicoli, Michele and Licheri, Marco and Maharana, Anshuman and Quevedo, Fernando},
  title         = {Logarithmic corrections and de Sitter vacua in string theory},
  journal       = {JHEP},
  volume        = {05},
  pages         = {001},
  year          = {2024},
  eprint        = {2401.16497},
  archivePrefix = {arXiv},
  primaryClass  = {hep-th}
}

@article{Gordon:2000hv,
  author        = {Gordon, Christopher and Wands, David and Bassett, Bruce A. and Maartens, Roy},
  title         = {Adiabatic and entropy perturbations from inflation},
  journal       = {Phys. Rev. D},
  volume        = {63},
  pages         = {023506},
  year          = {2000},
  doi           = {10.1103/PhysRevD.63.023506},
  eprint        = {astro-ph/0009131},
  archivePrefix = {arXiv}
}

@article{Langlois:2008qf,
  author={Langlois, David and Renaux-Petel, Sebastien},
  title={Perturbations in generalized multi-field inflation},
  journal={JCAP},
  volume={04},
  pages={017},
  year={2008},
  doi={10.1088/1475-7516/2008/04/017},
  eprint={0801.1085},
  archivePrefix={arXiv},
  primaryClass={hep-th}
}

@article{Tolley:2009fg,
  author={Tolley, Andrew J. and Wyman, Mark},
  title={The Gelaton Scenario: Equilateral non-Gaussianity from multi-field dynamics},
  journal={Phys. Rev. D},
  volume={81},
  pages={043502},
  year={2010},
  doi={10.1103/PhysRevD.81.043502},
  eprint={0910.1853},
  archivePrefix={arXiv},
  primaryClass={hep-th}
}

@article{Achucarro:2010da,
  author        = {Achucarro, Ana and Gong, Jinn-Ouk and Hardeman, Sjoerd and Palma, Gonzalo A. and Patil, Subodh P.},
  title         = {Features of heavy physics in the CMB power spectrum},
  journal       = {JCAP},
  volume        = {01},
  pages         = {030},
  year          = {2011},
  doi           = {10.1088/1475-7516/2011/01/030},
  eprint        = {1010.3693},
  archivePrefix = {arXiv},
  primaryClass  = {hep-ph}
}

@article{Peterson:2010np,
  author={Peterson, Courtney M. and Tegmark, Max},
  title={Testing multifield inflation: A geometric approach},
  journal={Phys. Rev. D},
  volume={83},
  pages={023522},
  year={2011},
  doi={10.1103/PhysRevD.83.023522},
  eprint={1005.4056},
  archivePrefix={arXiv},
  primaryClass={astro-ph.CO}
}

@article{Gong:2011uw,
  author={Gong, Jinn-Ouk and Tanaka, Takahiro},
  title={A covariant approach to general field space metric in multi-field inflation},
  journal={JCAP},
  volume={03},
  pages={015},
  year={2011},
  doi={10.1088/1475-7516/2011/03/015},
  eprint={1101.4809},
  archivePrefix={arXiv},
  primaryClass={astro-ph.CO}
}

@article{Achucarro:2012sm,
  author        = {Achucarro, Ana and Gong, Jinn-Ouk and Hardeman, Sjoerd and Palma, Gonzalo A. and Patil, Subodh P.},
  title         = {Effective theories of single field inflation when heavy fields matter},
  journal       = {JHEP},
  volume        = {05},
  pages         = {066},
  year          = {2012},
  doi           = {10.1007/JHEP05(2012)066},
  eprint        = {1201.6342},
  archivePrefix = {arXiv},
  primaryClass  = {hep-th}
}

@article{Cespedes:2012hu,
  author        = {Cespedes, Sebastian and Shiu, Gary and Tolley, Andrew J.},
  title         = {Inflationary non-Gaussianities in the most general second-order scalar-tensor theories},
  journal       = {JCAP},
  volume        = {02},
  pages         = {017},
  year          = {2013},
  doi           = {10.1088/1475-7516/2013/02/017},
  eprint        = {1201.6348},
  archivePrefix = {arXiv},
  primaryClass  = {hep-th}
}

@article{Burgess:2012dz,
  author={Burgess, C. P. and Horbatsch, M. and Patil, Subodh P.},
  title={Inflating in a Trough: Single-Field Effective Theory from Multiple-Field Curved Valleys},
  journal={JHEP},
  volume={01},
  pages={133},
  year={2013},
  doi={10.1007/JHEP01(2013)133},
  eprint={1209.5701},
  archivePrefix={arXiv},
  primaryClass={hep-th}
}

@misc{RenauxPetel:2015mga,
  author = {{S.~Renaux-Petel and K.~Turzynski}},
  title = {{Geometrical destabilization of inflation,}},
  year = {2016},
  note = {{Phys. Rev. Lett. \textbf{117}, 141301 (2016), arXiv:1510.01281 [astro-ph.CO].}}
}

@article{Cheung:2007st,
  author = {Cheung, Clifford and Creminelli, Paolo and Fitzpatrick, A. Liam and Kaplan, Jared and Senatore, Leonardo},
  title = {The effective field theory of inflation},
  journal = {JHEP},
  volume = {03},
  pages = {014},
  year = {2008},
  eprint = {0709.0293},
  archivePrefix = {arXiv},
  primaryClass = {hep-th}
}

@article{Weinberg:2008hq,
  author = {Weinberg, Steven},
  title = {Effective field theory for inflation},
  journal = {Phys. Rev. D},
  volume = {77},
  pages = {123541},
  year = {2008},
  eprint = {0804.4291},
  archivePrefix = {arXiv},
  primaryClass = {hep-th}
}

@article{Senatore:2010wk,
  author = {Senatore, Leonardo and Zaldarriaga, Matias},
  title = {The effective field theory of multifield inflation},
  journal = {JHEP},
  volume = {04},
  pages = {024},
  year = {2012},
  eprint = {1009.2093},
  archivePrefix = {arXiv},
  primaryClass = {hep-th}
}

@article{Shiu:2011qw,
  author = {Shiu, Gary and Xu, Jiajun},
  title = {Effective field theory and decoupling in multi-field inflation},
  journal = {Phys. Rev. D},
  volume = {84},
  pages = {103509},
  year = {2011},
  eprint = {1108.0981},
  archivePrefix = {arXiv},
  primaryClass = {hep-th}
}

@article{Komatsu:2009kd,
  author = {Komatsu, Eiichiro},
  title = {Hunting for primordial non-Gaussianity in the cosmic microwave background},
  journal = {Class. Quant. Grav.},
  volume = {27},
  pages = {124010},
  year = {2010},
  eprint = {1003.6097},
  archivePrefix = {arXiv},
  primaryClass = {astro-ph.CO}
}

@article{Chen:2010xka,
  author = {Chen, Xingang},
  title = {Primordial non-Gaussianities from inflation models},
  journal = {Adv. Astron.},
  volume = {2010},
  pages = {638979},
  year = {2010},
  eprint = {1002.1416},
  archivePrefix = {arXiv},
  primaryClass = {astro-ph.CO}
}

@article{Bousso:2000xa,
  author={Bousso, Raphael and Polchinski, Joseph},
  title={Quantization of four form fluxes and dynamical neutralization of the cosmological constant},
  journal={JHEP},
  volume={06},
  pages={006},
  year={2000},
  doi={10.1088/1126-6708/2000/06/006},
  eprint={hep-th/0004134},
  archivePrefix={arXiv}
}

@article{Douglas:2003um,
  author = {Douglas, Michael R.},
  title = {The statistics of string/M theory vacua},
  journal = {JHEP},
  volume = {05},
  pages = {046},
  year = {2003},
  eprint = {hep-th/0303194},
  archivePrefix = {arXiv}
}

@article{Ashok:2003gk,
  author = {Ashok, Sujay K. and Douglas, Michael R.},
  title = {Counting flux vacua},
  journal = {JHEP},
  volume = {01},
  pages = {060},
  year = {2004},
  eprint = {hep-th/0307049},
  archivePrefix = {arXiv}
}

@article{Denef:2004dm,
  author={Denef, Frederik and Douglas, Michael R.},
  title={Distributions of flux vacua},
  journal={JHEP},
  volume={05},
  pages={072},
  year={2004},
  doi={10.1088/1126-6708/2004/05/072},
  eprint={hep-th/0404116},
  archivePrefix={arXiv}
}

@article{Denef:2005mm,
  author = {Denef, Frederik and Douglas, Michael R.},
  title = {Distributions of nonsupersymmetric flux vacua},
  journal = {JHEP},
  volume = {03},
  pages = {061},
  year = {2005},
  eprint = {hep-th/0411183},
  archivePrefix = {arXiv}
}

@article{Douglas:2006es,
  author={Douglas, Michael R. and Kachru, Shamit},
  title={Flux compactification},
  journal={Rev. Mod. Phys.},
  volume={79},
  pages={733--796},
  year={2007},
  doi={10.1103/RevModPhys.79.733},
  eprint={hep-th/0610102},
  archivePrefix={arXiv}
}

@article{Saltman:2004sn,
  author={Saltman, Alexander and Silverstein, Eva},
  title={The Scaling of the no scale potential and de Sitter model building},
  journal={JHEP},
  volume={11},
  pages={066},
  year={2004},
  doi={10.1088/1126-6708/2004/11/066},
  eprint={hep-th/0402135},
  archivePrefix={arXiv}
}

@article{Blumenhagen:2006ci,
  author={Blumenhagen, Ralph and Kors, Boris and Lust, Dieter and Stieberger, Stephan},
  title={Four-dimensional String Compactifications with D-Branes, Orientifolds and Fluxes},
  journal={Phys. Rept.},
  volume={445},
  pages={1--193},
  year={2007},
  doi={10.1016/j.physrep.2007.04.003},
  eprint={hep-th/0610327},
  archivePrefix={arXiv}
}

@article{Grimm:2024exactFlux,
  author        = {Grimm, Thomas W. and van de Heisteeg, Damian},
  title         = {Exact Flux Vacua, Symmetries, and the Structure of the Landscape},
  journal       = {JHEP},
  volume        = {01},
  pages         = {005},
  year          = {2025},
  doi           = {10.1007/JHEP01(2025)005},
  eprint        = {2404.12422},
  archivePrefix = {arXiv},
  primaryClass  = {hep-th}
}

@article{McAllister:2024dS,
  author        = {McAllister, Liam and Moritz, Jakob and Nally, Richard and Schachner, Andreas},
  title         = {Candidate de Sitter Vacua},
  journal       = {Phys. Rev. D},
  volume        = {111},
  number        = {8},
  pages         = {086015},
  year          = {2025},
  doi           = {10.1103/PhysRevD.111.086015},
  eprint        = {2406.13751},
  archivePrefix = {arXiv},
  primaryClass  = {hep-th}
}

@misc{Schachner:2025overview,
  author        = {Schachner, Andreas},
  title         = {Brief overview of Candidate de Sitter Vacua},
  year          = {2025},
  eprint        = {2505.00149},
  archivePrefix = {arXiv},
  primaryClass  = {hep-th},
  note          = {Contribution to the proceedings of CORFU2024}
}

@article{Kachru:2002gs,
  author={Kachru, Shamit and Pearson, John and Verlinde, Herman L.},
  title={Brane/flux annihilation and the string dual of a nonsupersymmetric field theory},
  journal={JHEP},
  volume={06},
  pages={021},
  year={2002},
  doi={10.1088/1126-6708/2002/06/021},
  eprint={hep-th/0112197},
  archivePrefix={arXiv}
}

@article{Bena:2009xk,
  author = {Bena, Iosif and Grana, Mariana and Halmagyi, Nick},
  title = {On the existence of meta-stable vacua in Klebanov-Strassler},
  journal = {JHEP},
  volume = {09},
  pages = {087},
  year = {2010},
  eprint = {0912.3519},
  archivePrefix = {arXiv},
  primaryClass = {hep-th}
}

@article{Bena:2010ze,
  author = {Bena, Iosif and Giecold, Gregory and Grana, Mariana and Halmagyi, Nick},
  title = {On the inflaton potential from antibranes in warped throats},
  journal = {JHEP},
  volume = {07},
  pages = {140},
  year = {2011},
  eprint = {1011.2626},
  archivePrefix = {arXiv},
  primaryClass = {hep-th}
}

@article{Bena:2012vz,
  author = {Bena, Iosif and Grana, Mariana and Halmagyi, Nick and Massai, Stefano},
  title = {On metastable vacua and the warped deformed conifold},
  journal = {JHEP},
  volume = {06},
  pages = {044},
  year = {2013},
  eprint = {1212.4828},
  archivePrefix = {arXiv},
  primaryClass = {hep-th}
}

@article{Blaback:2012nf,
  author = {Blaback, Johan and Danielsson, Ulf H. and Van Riet, Thomas},
  title = {Resolving anti-brane singularities through time-dependence},
  journal = {JHEP},
  volume = {02},
  pages = {061},
  year = {2013},
  eprint = {1202.1132},
  archivePrefix = {arXiv},
  primaryClass = {hep-th}
}

@article{Danielsson:2014yga,
  author = {Danielsson, Ulf H. and Gautason, Fridrik Freyr and Van Riet, Thomas},
  title = {Unstoppable brane-flux decay of D6 branes},
  journal = {JHEP},
  volume = {03},
  pages = {141},
  year = {2017},
  eprint = {1609.06529},
  archivePrefix = {arXiv},
  primaryClass = {hep-th}
}

@article{Michel:2014lva,
  author = {Michel, Ben and Mintun, Eric and Polchinski, Joseph and Puhm, Andrea and Saad, Philip},
  title = {Remarks on brane and antibrane dynamics},
  journal = {JHEP},
  volume = {09},
  pages = {021},
  year = {2015},
  eprint = {1412.5702},
  archivePrefix = {arXiv},
  primaryClass = {hep-th}
}

@article{CohenMaldonado:2015ssa,
  author = {Cohen-Maldonado, Diego and Diaz, Dimitri Z. and Van Riet, Thomas and Vercnocke, Bert},
  title = {Observations on fluxes near anti-branes},
  journal = {JHEP},
  volume = {01},
  pages = {126},
  year = {2016},
  eprint = {1507.01022},
  archivePrefix = {arXiv},
  primaryClass = {hep-th}
}

@article{Polchinski:2015bea,
  author = {Polchinski, Joseph},
  title = {Brane/antibrane dynamics and KKLT stability},
  journal = {arXiv e-prints},
  pages = {arXiv:1509.05710},
  year = {2015},
  eprint = {1509.05710},
  archivePrefix = {arXiv},
  primaryClass = {hep-th}
}

@article{Moritz:2017xto,
  author = {Moritz, Jakob and Retolaza, Ander and Westphal, Alexander},
  title = {Toward de Sitter space from ten dimensions},
  journal = {Phys. Rev. D},
  volume = {97},
  pages = {046010},
  year = {2018},
  eprint = {1707.08678},
  archivePrefix = {arXiv},
  primaryClass = {hep-th}
}

@article{Gautason:2018gln,
  author = {Gautason, Fridrik Freyr and Junghans, Daniel and Zagermann, Marco},
  title = {On cosmological constants from alpha-prime corrections},
  journal = {JHEP},
  volume = {06},
  pages = {029},
  year = {2012},
  eprint = {1204.0807},
  archivePrefix = {arXiv},
  primaryClass = {hep-th}
}

@article{Junghans:2018gdb,
  author = {Junghans, Daniel},
  title = {Weakly coupled de Sitter vacua with fluxes and the swampland},
  journal = {JHEP},
  volume = {03},
  pages = {150},
  year = {2019},
  eprint = {1811.06990},
  archivePrefix = {arXiv},
  primaryClass = {hep-th}
}

@article{Sethi:2017phn,
  author = {Sethi, Savdeep},
  title = {Supersymmetry breaking by fluxes},
  journal = {JHEP},
  volume = {10},
  pages = {022},
  year = {2018},
  eprint = {1709.03554},
  archivePrefix = {arXiv},
  primaryClass = {hep-th}
}

@article{Kachru:2018aqn,
  author = {Kachru, Shamit and Trivedi, Sandip P.},
  title = {A comment on effective field theories of flux vacua},
  journal = {Fortsch. Phys.},
  volume = {67},
  pages = {1800086},
  year = {2019},
  eprint = {1808.08971},
  archivePrefix = {arXiv},
  primaryClass = {hep-th}
}

\end{document}